\documentclass[aps, pra, twocolumn, floatfix, amssymb, nobibnotes, altaffilletter, superscriptaddress, raggedbottom]{revtex4-1}

\usepackage{amsmath} 
\usepackage{color}
\usepackage{graphicx}
\usepackage{SIunits}
\usepackage{bm}
\usepackage{multirow,diagbox}
\usepackage{balance}

\newcommand{\mk}[1]{\underline{#1}}
\newcommand{\beginsupplement}{%
    \setcounter{table}{0}
    \renewcommand{\thetable}{A\arabic{table}}%
    \setcounter{figure}{0}
    \renewcommand{\thefigure}{A\arabic{figure}}%
}

\begin{document}

\title{Experimental 3D Coherent Diffractive Imaging from photon-sparse random projections}

\author{K. Giewekemeyer}
\email[ ]{klaus.giewekemeyer@xfel.eu}
\affiliation{European XFEL GmbH, Holzkoppel 4, 22869 Schenefeld, Germany}
\author{A. Aquila}
\altaffiliation[present address: ]{SLAC National Accelerator Laboratory, 2575 Sand Hill Road, Menlo Park (CA) 94025, USA}
\affiliation{European XFEL GmbH, Holzkoppel 4, 22869 Schenefeld, Germany}
\author{N.D. Loh}
\affiliation{Centre for Bio-imaging Sciences, National University of Singapore, 14 Science Drive 4, 117557 Singapore}
\affiliation{Department of Physics, National University of Singapore, 2 Science Drive 3, 117551 Singapore}
\affiliation{Department of Biological Sciences, National University of Singapore, 14 Science Drive 4, 117557 Singapore}
\author{Y. Chushkin}
\affiliation{ESRF - The European Synchrotron, 71 Avenue des Martyrs, 38000 Grenoble, France}
\author{K.S. Shanks}
\affiliation{Laboratory for Atomic and Solid State Physics, Cornell University, Ithaca (NY) 14853, USA}
\author{J. Weiss}
\affiliation{Laboratory for Atomic and Solid State Physics, Cornell University, Ithaca (NY) 14853, USA}
\author{M.W. Tate}
\affiliation{Laboratory for Atomic and Solid State Physics, Cornell University, Ithaca (NY) 14853, USA}
\author{H.T. Philipp}
\affiliation{Laboratory for Atomic and Solid State Physics, Cornell University, Ithaca (NY) 14853, USA}
\author{S. Stern}
\affiliation{European XFEL GmbH, Holzkoppel 4, 22869 Schenefeld, Germany}
\affiliation{Center for Free-Electron Laser Science, Deutsches Elektronen-Synchrotron, 22607 Hamburg, Germany}
\author{P. Vagovic}
\affiliation{European XFEL GmbH, Holzkoppel 4, 22869 Schenefeld, Germany}
\affiliation{Center for Free-Electron Laser Science, Deutsches Elektronen-Synchrotron, 22607 Hamburg, Germany}
\author{M. Mehrjoo}
\altaffiliation[present address: ]{Deutsches Elektronen-Synchrotron, Notkestraße 85, 22607 Hamburg, Germany}
\affiliation{European XFEL GmbH, Holzkoppel 4, 22869 Schenefeld, Germany}
\author{C. Teo}
\affiliation{Centre for Bio-imaging Sciences, National University of Singapore, 14 Science Drive 4, 117557 Singapore}
\affiliation{Department of Biological Sciences, National University of Singapore, 14 Science Drive 4, 117557 Singapore}
\author{M. Barthelmess}
\affiliation{Center for Free-Electron Laser Science, Deutsches Elektronen-Synchrotron, 22607 Hamburg, Germany}
\author{F. Zontone}
\affiliation{ESRF - The European Synchrotron, 71 Avenue des Martyrs, 38000 Grenoble, France}
\author{C. Chang}
\affiliation{SLAC National Accelerator Laboratory, 2575 Sand Hill Road, Menlo Park (CA) 94025, USA}
\author{Richard C. Tiberio}
\affiliation{Stanford Nano Shared Facilities, Stanford University, 348 Via Pueblo, Stanford (CA) 94305, USA}
  \author{A. Sakdinawat}
\affiliation{SLAC National Accelerator Laboratory, 2575 Sand Hill Road, Menlo Park (CA) 94025, USA}
\author{G.J. Williams}
\altaffiliation[present address: ]{NSLS-II, Brookhaven National Laboratory, P.O. Box 5000, Upton (NY) 11973, USA}
\affiliation{SLAC National Accelerator Laboratory, 2575 Sand Hill Road, Menlo Park (CA) 94025, USA}
\author{S.M. Gruner}
\affiliation{Laboratory for Atomic and Solid State Physics, Cornell University, Ithaca (NY) 14853, USA}
\affiliation{Cornell High Energy Synchrotron Source (CHESS), Cornell University, Ithaca (NY) 14853, USA}
\affiliation{Kavli Institute at Cornell for Nanoscale Science, Cornell University, Ithaca (NY) 14853, USA}
\author{A.P. Mancuso}
\affiliation{European XFEL GmbH, Holzkoppel 4, 22869 Schenefeld, Germany}

\begin{abstract}
The routine atomic-resolution structure determination of single particles is expected to have profound implications for probing the structure-function relationship in systems ranging from energy materials to biological molecules. Extremely-bright, ultrashort-pulse X-ray sources---X-ray Free Electron Lasers (XFELs)---provide X-rays that can be used to probe ensembles of nearly identical nano-scale particles. When combined with coherent diffractive imaging, these objects can be imaged; however, as the resolution of the images approaches the atomic scale, the measured data are increasingly difficult to obtain and, during an X-ray pulse, the number of photons incident on the two-dimensional detector is much smaller than the number of pixels. This latter concern, the signal ``sparsity,'' materially impedes the application of the method. We demonstrate an experimental analog using a synchrotron X-ray source that yields signal levels comparable to those expected from single biomolecules illuminated by focused XFEL pulses. The analog experiment provides an invaluable cross-check on the fidelity of the reconstructed data that is not available during XFEL experiments. We establish---using this experimental data---that a sparsity of order $1.3\times10^{-3}$ photons per pixel per frame can be overcome, lending vital insight to the solution of the atomic-resolution XFEL single particle imaging problem by experimentally demonstrating 3D coherent diffractive imaging from photon-sparse random projections.
\end{abstract}

\maketitle

\section{Introduction}

A major motivation for advancing Coherent Diffractive Imaging (CDI) using X-rays has always been its potential application to the imaging of individual nano-scale objects. A specific case concerns biological macromolecules, where structure can be determined without the need for crystallization\cite{neutze_potential_2000,shenoy_lcls_2003}, representing an early potential application of X-ray free electron lasers (XFELs). The high peak flux, of order \unit{\power{10}{12}}{photons/pulse}, and the shorter than \unit{100}{\femto\second}-scale pulse duration are the prerequisites for generating diffraction signal from a single macromolecule, which is destroyed by the Coulomb explosion of the sample \cite{neutze_potential_2000} resulting from ionization during the measurement. This scheme of `diffraction-before-destruction' is routinely used for Serial Femtosecond Crystallography \cite{chapman_femtosecond_2011,schlichting_serial_2015} and has been demonstrated for single biological particles as small as $\unit{45}{\nano\metre}$ in diameter \cite{daurer_experimental_2017}. 

To obtain 3D structural information from such single particles, serial diffraction data from many identical or nearly identical objects has to be measured with sufficient orientational variation. Usually, this is achieved by randomly injecting particles into the FEL beam, relying on the statistical coincidence of a single particle to be hit by an FEL pulse \cite{barty_single_2016,spence_xfels_2017}. Due to the random nature of this process, the sample's orientation for each diffraction pattern is generally unknown and has to be recovered \textit{a posteriori} in order to build up a continuous 3D diffraction volume in reciprocal-space. This can then be inverted by iterative phase retrieval \cite{marchesini_invited_2007} into a real-space electron density distribution, the last step of Single Particle CDI.

A first complete demonstration of the method was provided by the 3D structure determination of the Giant Mimivirus, approximately \unit{450}{\nano\metre} in diameter, to a resolution \footnote{Unless otherwise noted, we define resolution as the crystallographic (full-period) resolution, in contrast to the half-period resolution which is also often used in the literature on Coherent Diffractive Imaging.} of \unit{125}{\nano\metre} \cite{ekeberg_three-dimensional_2015}. Very recently, a step towards much smaller viruses---the Rice Dwarf Virus (RDV) and bacteriophage PR772, both with a diameter around \unit{70}{\nano\metre}---has been made, resulting in images at a resolution slightly above \unit{10}{\nano\metre} \cite{kurta_correlations_2017} and, more recently, for PR772, slighly below \unit{10}{\nano\metre} \cite{rose_single-particle_2018}. Those data were a result of the Single-Particle-Imaging Initiative at the Linac Coherent Light Source \cite{aquila_linac_2015,munke_coherent_2016,reddy_coherent_2017}. This also resulted in the collection of a few hundred high-resolution diffraction frames from RDV at a photon energy of \unit{7}{\kilo\electronvolt}, showing that useful diffraction signal can be collected at \unit{5.9}{\angstrom} resolution from single hits \cite{munke_coherent_2016}. 

Reconstructing biological macromolecules to \unit{3}{\angstrom} or better resolution has previously been set as the ultimate goal of Single Particle CDI \cite{aquila_linac_2015}. When approaching the molecular-size scale, the diffracted signal becomes very sparse, with a typical pattern containing less than a few hundred diffracted photons \cite{yoon_comprehensive_2016,fortmann-grote_start--end_2017}. For example, a protein of at least ca.\ \unit{10}{\nano\metre} diameter is required to scatter, on average, 50 photons outside the central speckle, at a photon energy of \unit{8}{\kilo\electronvolt} in a nano-scale FEL focus (for further details, see below). In this case, hundreds of thousands of diffraction patterns have to be collected to build up the 3D reciprocal-space intensity, i.e., to assemble an invertible dataset \cite{loh_reconstruction_2009}. To date, no such experimental dataset exists and considerable method development is still required towards the realization of Single Particle CDI as an independent method of macromolecular structure determination.

An important branch of this method development addresses the problem of orientation recovery in the case of very weak diffraction which is often not only sparse, but also contaminated by background signal, originating from sources such as the instrument or the particle beam. In recent years, several methods for orientation recovery have been devised \cite{huldt_diffraction_2003,loh_reconstruction_2009,fung_structure_2009,bortel_common_2011,tegze_atomic_2012,giannakis_symmetries_2012,yefanov_orientation_2013,kassemeyer_optimal_2013,zhou_multiple-common-lines_2014,donatelli_iterative_2015,flamant_expansion-maximization-compression_2016,donatelli_reconstruction_2017,nakano_three-dimensional_2017,ardenne_structure_2018} and also applied to experimental Single Particle FEL \cite{loh_cryptotomography:_2010,kassemeyer_optimal_2013,ekeberg_three-dimensional_2015,kurta_correlations_2017,lundholm_considerations_2018,ardenne_structure_2018} or similar data \cite{nakano_three-dimensional_2017}. 

However, the important case of sparse diffraction from a 3D object has only been solved experimentally in a setting different from the classical CDI problem. For example, one of the methods of orientation recovery--a statistical technique based on expectation maximization, the Expand-Maximize-Compress (EMC) algorithm \cite{loh_reconstruction_2009,ayyer_dragonfly:_2016}--has been applied successfully to real-space sparse radiographic data, in two \cite{philipp_solving_2012} and three dimensions \cite{ayyer_real-space_2014}, to sparse crystallographic data limited to one \cite{wierman_protein_2016} and two rotation axes \cite{lan_reconstructing_2017}, and very recently also to synchrotron-based serial protein crystallographic data for random crystal orientations \cite{lan_solving_2018}.  

Here, we demonstrate 3D CDI from sparse random projections in the same geometry as that used for FEL-based Single Particle Imaging experiments. Using a synchrotron beam on a micron-scale sample, we show that, with as few as $50$ scattered photons per diffraction pattern, and without explicit knowledge of the sample's orientation for a given data frame, it is possible to robustly reconstruct the scattering distribution of the sample in reciprocal-space and to invert this diffraction volume into a high-resolution 3D electron density distribution. We show that, despite strong sparsity in the data, it is possible to reconstruct a sample to a complexity of more than $40$ resolution elements \footnote{For easy comparison, the term `resolution element' is defined in analogy to \cite{loh_reconstruction_2009} and \cite{ayyer_dragonfly:_2016}, i.e., two resolution elements equal the full-period (crystallographic) resolution. While in simulations it is often defined by the edge of the detector, we define it in terms of the resolution as obtained after phasing.} within the largest diameter of the sample.

\section{Experiment}

The experiment was performed at the undulator beamline ID10, end station EH2, of the European Synchrotron Radiation Facility (ESRF) \cite{chushkin_three-dimensional_2014}. The photon energy was set to $8.1$~keV using a water-cooled Si(111) pseudo-channel-cut monochromator with an intrinsic energy resolution of $\Delta E/E \simeq 1.4\cdot10^{-4}$. The sample---a solid gold object with a largest diagonal length of about \unit{1.1}{\micro\metre}, fabricated by electroplating and supported by a silicon-nitride membrane---was placed into the beam on a high-precision tomographic stage at a distance of \unit{4.0}{\metre} from the detector. Defined by several sets of slits \cite{chushkin_three-dimensional_2014}, the lateral beam size at the sample plane was approximately \unit{10}{\micro\metre} $\times$ \unit{10}{\micro\metre}. A schematic of the setup is shown in Fig.\ 1. For further details on all major aspects of our data treatment and algorithms employed in the reciprocal- and real-space reconstruction of our data, we refer the reader to the appendix.

\begin{figure}
\centering
\includegraphics[width=0.95\columnwidth]{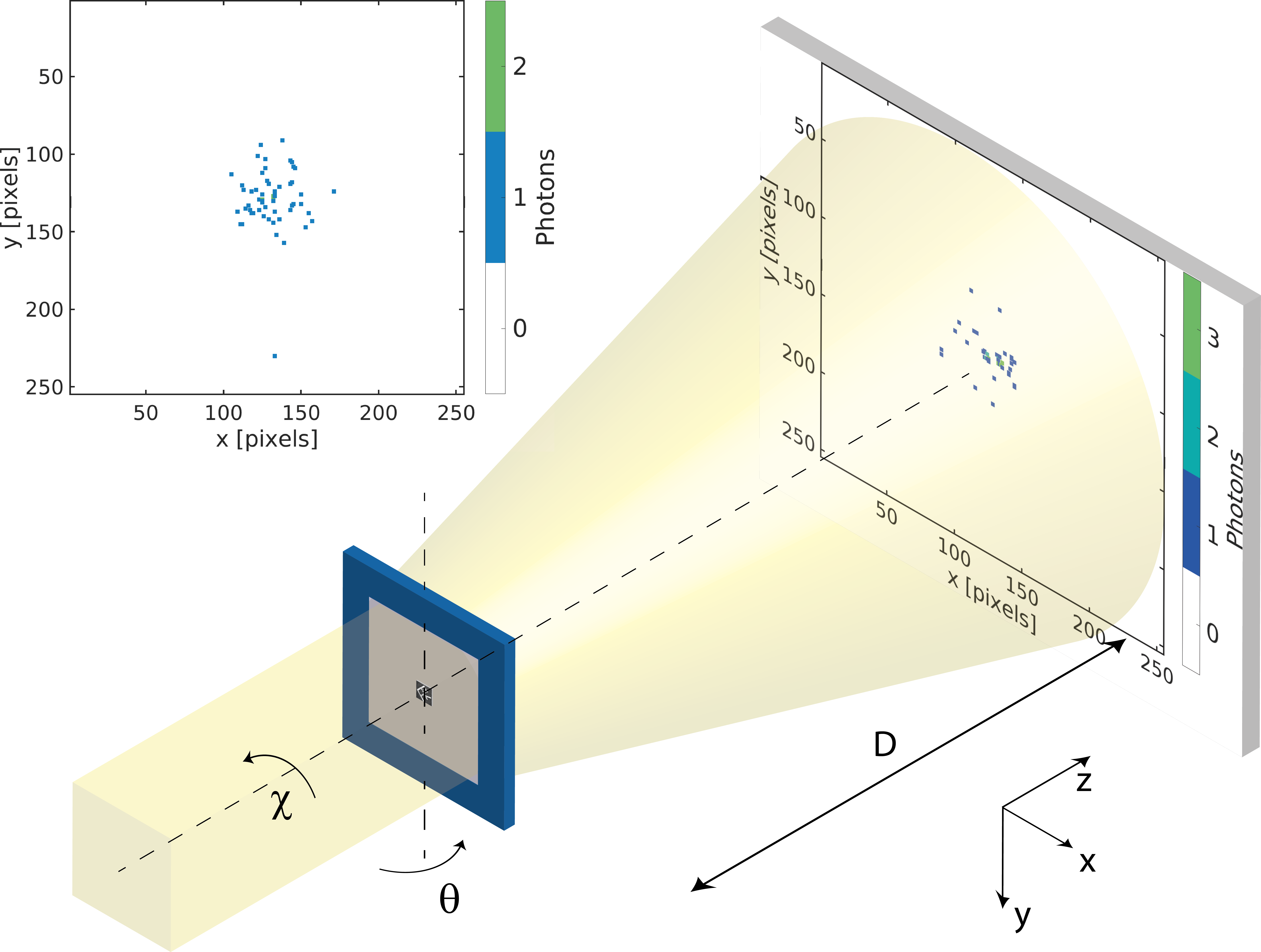}
\caption{Schematic of the experiment. The sample, a gold nanostructure, supported on a silicon nitride membrane, was rotated about the $y$-axis by angle $\theta$ to obtain diffraction patterns at different orientations with respect to the optical axis, $z$. A first rotation series about the $y$-axis was followed by an in-plane rotation of the sample about $z$ (angle $\chi$) and a subsequent second rotation series about $y$. The beam attenuation and illumination time were adjusted, so that each data frame contains only about $50$ scattered photons. An example of a single diffraction pattern is shown in the inset on the upper left.}
\label{fg:setup}
\end{figure}

For data collection, the Mixed-Mode Pixel Array Detector (MM-PAD) was used. It is a wide dynamic range integrating detector developed at Cornell University that is capable of collecting data at a kHz frame rate with a high signal-to-noise ratio, ranging from one x-ray photon/pixel to a maximum rate exceeding $10^8$~photons/pixel/s at \unit{8}{\kilo\electronvolt} photon energy \cite{tate_medium-format_2013}. The beam was attenuated using a polished single-crystal Si attenuator with a transmission of ca.\ $0.3$. As a result, the overall flux reaching the detector was approximately $10^8$~photons/s.

The dataset analyzed here comprises diffraction patterns from 227 unique 3D orientations. With 2000 collected frames per orientation, this amounts to a total of $M_{data} = 454\,000$ data frames. Each frame was collected with an illumination time of $25$~ms. The orientations were obtained from two independent tomographic series (see Fig.\ \ref{fg:setup}.). During each of these, only $\theta$, the angle about the tomographic rotation axis $y$, was varied.

\section{Analysis}

\subsection{Detector calibration and further data treatment}

As the collected data were very sparse, calibration of the raw detector data is an essential step of the analysis. The applied calibration procedure consisted of several steps. A binary mask was used to reject all pixels that were inactive or were within gaps between detector modules. Second, the dark signal was subtracted for each pixel. The gain of the detector was determined from the histogram of pixel values over many frames within a region of interest where the maximum intensity/pixel is only a few photons per frame. The resulting histogram shows discrete peaks corresponding to zero, one, two, etc. photons per pixel. In terms of Analog-to-Digital units (ADU), a gain of $11.1$ ADU/photon at $8.1$ keV was determined from the spacing of these peaks. This gain, together with the width of these peaks, yields a Signal-to-Noise ratio (SNR) of $5.2$ for single photons. For transformation of ADUs into single photons, a threshold energy of $E_t = E_{\gamma} - \textrm{HWHM}_n = 0.77E_{\gamma}$ was used, with the single-photon energy $E_\gamma$ and the noise peak Half Width at Half Maximum ($\textrm{HWHM}_n$). With around $40\,000$ active detector pixels in a region of interest (ROI) of $255\times255$ pixels around the beam center, this leads to a false-positive probability $P(1|0)\approx3\times 10^{-5}$ or between 1 and 2 events per frame. At the sparsity level of the data and with more than $400\,000$ frames comprising the complete dataset, it was found that cosmic radiation made up a considerable contribution to the total recorded intensity. We therefore applied a threshold-based removal procedure. The sparsity level of the single frames is reflected by an average of $49.3$ scattered photons per frame \footnote{The median amounts to $48$ and the standard deviation to $9.6$.}, after masking out pixels dominated by empty-beam scattering. Due to the low level of background scattering from the instrument---e.g., optical components or apertures---no background or `empty-beam' subtraction was performed on the data.

\subsection{Orientation determination}

The goal of orientation determination is to obtain the 3D reciprocal-space intensity $W(\mathbf{q})$ that is proportional to the modulus-squared Fourier transform of the 3D electron density of the sample. Here, $\bf{q}$ denotes the 3D Cartesian reciprocal-space coordinate. To reconstruct $W(\mathbf{q})$, the EMC algorithm correlates each data frame, $K_d$ ($d = 1,\dots,M_\mathrm{data}$), with tomographic slices $W_j$ ($j = 1,\dots,M_\mathrm{rot}$) of $W$, corresponding to $M_\mathrm{rot}$ possible sample orientations, based on the current iterate of our model of $W(\mathbf{q})$. Each iteration comprises expanding the current model $W$ into slices $W_{j}$, an update $W_j \rightarrow W_j'$ by maximizing a log-likelihood function $Q(W')$, and compressing slices $W_j'$ into a new 3D model $W'(\mathbf{q})$. The update itself consists of forming the weighted sum $W'_{ij} = \sum_{d=1}^{M_\mathrm{data}} P_{jd}(W)K_{id} / \sum_{d=1}^{M_\mathrm{data}}P_{jd}(W)$. Index $i$ specifies a pixel and $P_{jd}(W)$ is the probability of frame $K_d$ having been collected at orientation $j$, based on the slice $W_j$. Each orientation $j$ of the sample is represented by a unit quaternion, $\mk{q}_j$. For the present experiment, the set $\{\mk{q}_j\}$ of possible orientations was derived from the experimental setup and procedure (for details, see the appendix). As a result, the optimized set $\{\mk{q}_j\}$ of orientations could be used to generate a Fourier intensity $W_\mathrm{ref}(\mathbf{q})$, assembled using full knowledge of orientations, as a reference for the EMC-based intensity reconstruction. EMC  received the same set of orientations as an input together with all data frames, but without explicit knowledge any frame's orientation. As a further input, a geometry file was included that contained the reciprocal-space coordinate of each detector pixel and a 3D binary mask, $S$, identifying those voxels in the cubic domain of $W(\mathbf{q})$ that are to be excluded from the analysis process, as e.g., they are never reached by any Ewald sphere slice. As $S$ is non-symmetric with respect to $\mathbf{q}=0$, the Friedel symmetrization step included in EMC \cite{loh_reconstruction_2009} was modified accordingly. To account for the fact that some pixels in each frame $K_d$ contain a large fraction of sample scattering, but still have a significant contribution of `parasitic' or beamline scatter, we defined a binary mask on the detector ROI to identify those pixels to be included in the update rule $W \rightarrow W'$, but not into the calculation of probabilities $P_{jd}(W)$ \cite{ayyer_dragonfly:_2016}.

As EMC was always initiated with a random intensity distribution, independent runs of the algorithm show some statistical variation. To reduce the associated uncertainty, the EMC algorithm was run 20 times for 500 iterations, followed by an averaging procedure similar to that described in \cite{yoon_comprehensive_2016}. A small fraction of the ensemble, 2 out of 20 reconstructions, exhibited artifacts due to localized over-weighting of certain orientations. These could be automatically discarded by rejecting highly non-homogeneous distributions of orientations. Within the remaining results, two main classes could be observed which are related by an overall rotation of about 180 degrees around an axis close to one of the coordinate axes. This is in accordance with a previous study for an isotropic orientational distribution \cite{yoon_comprehensive_2016}. After manual attribution to one of the two classes the results were averaged and their relation was verified by orientational registration. The final averaged 3D reciprocal-space volume $\langle W(q) \rangle$ was obtained as an average of 13 individual results in the same orientation as $W_\mathrm{ref}(\mathrm{q})$.

\section{Results}

\begin{figure}
\centering
\includegraphics[width=0.9\columnwidth]{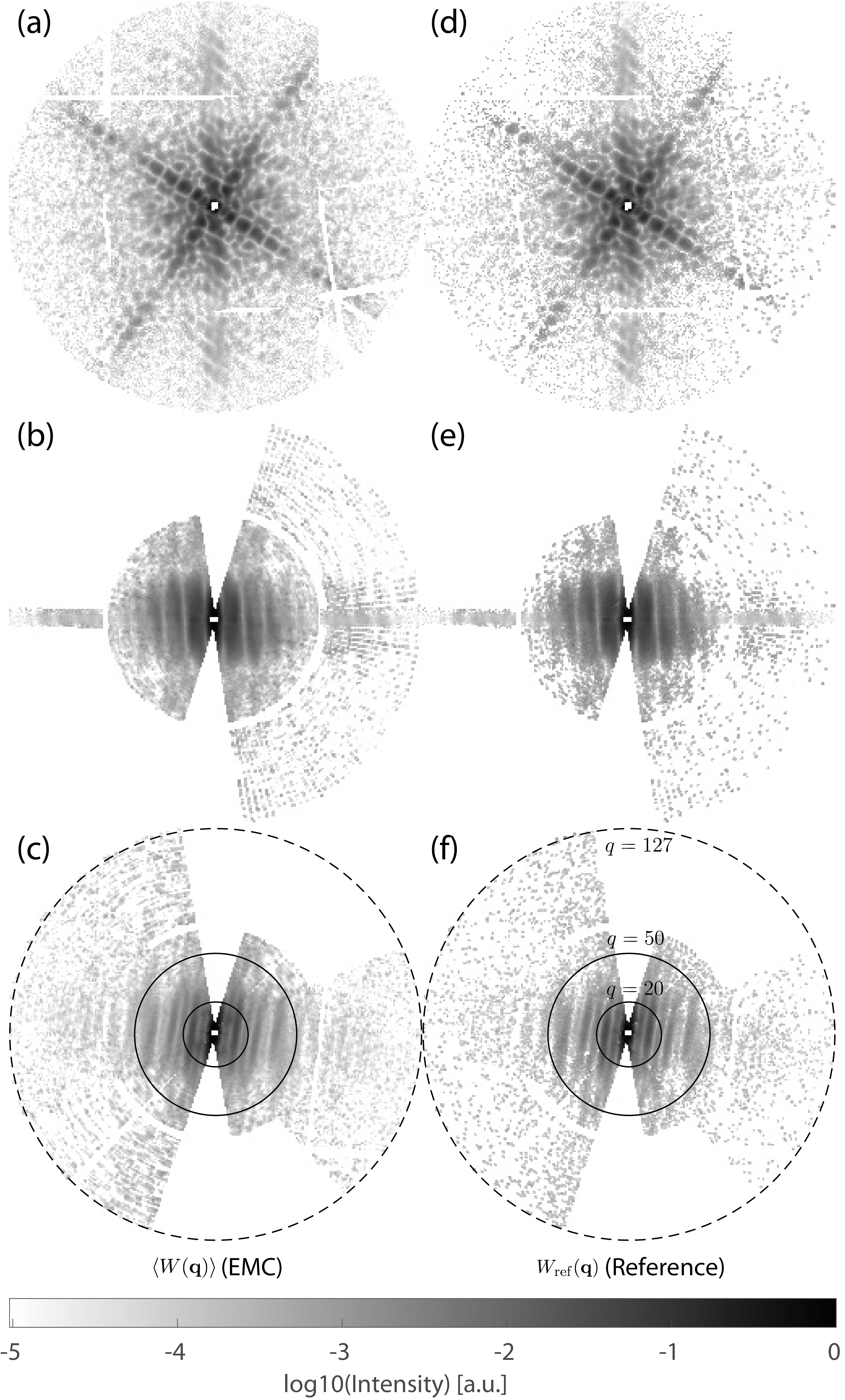}
\caption{Orthogonal slices through the EMC-reconstructed (a-c) and the reference (d-f) 3D diffraction volume. The EMC-reconstructed diffraction volume $\langle W(\mathbf{q})\rangle$ results from averaging the results of 13 independent EMC runs, each starting with a random intensity distribution. The reference diffraction volume $W_\mathrm{ref}(\mathbf{q})$ was constructed based on the known orientations of the sample for each frame during the measurement. Dashed circles in subfigures (c) and (f) indicate a radius of 127 voxels, whereas the solid circles indicate radii of 20 and 50, respectively. All slices are drawn on the same scale with dimensionless lateral coordinates in units of $k \Delta X /D$.}
\label{fg:slices_rec}
\end{figure}

\begin{figure}
\centering
\includegraphics[width=0.95\linewidth]{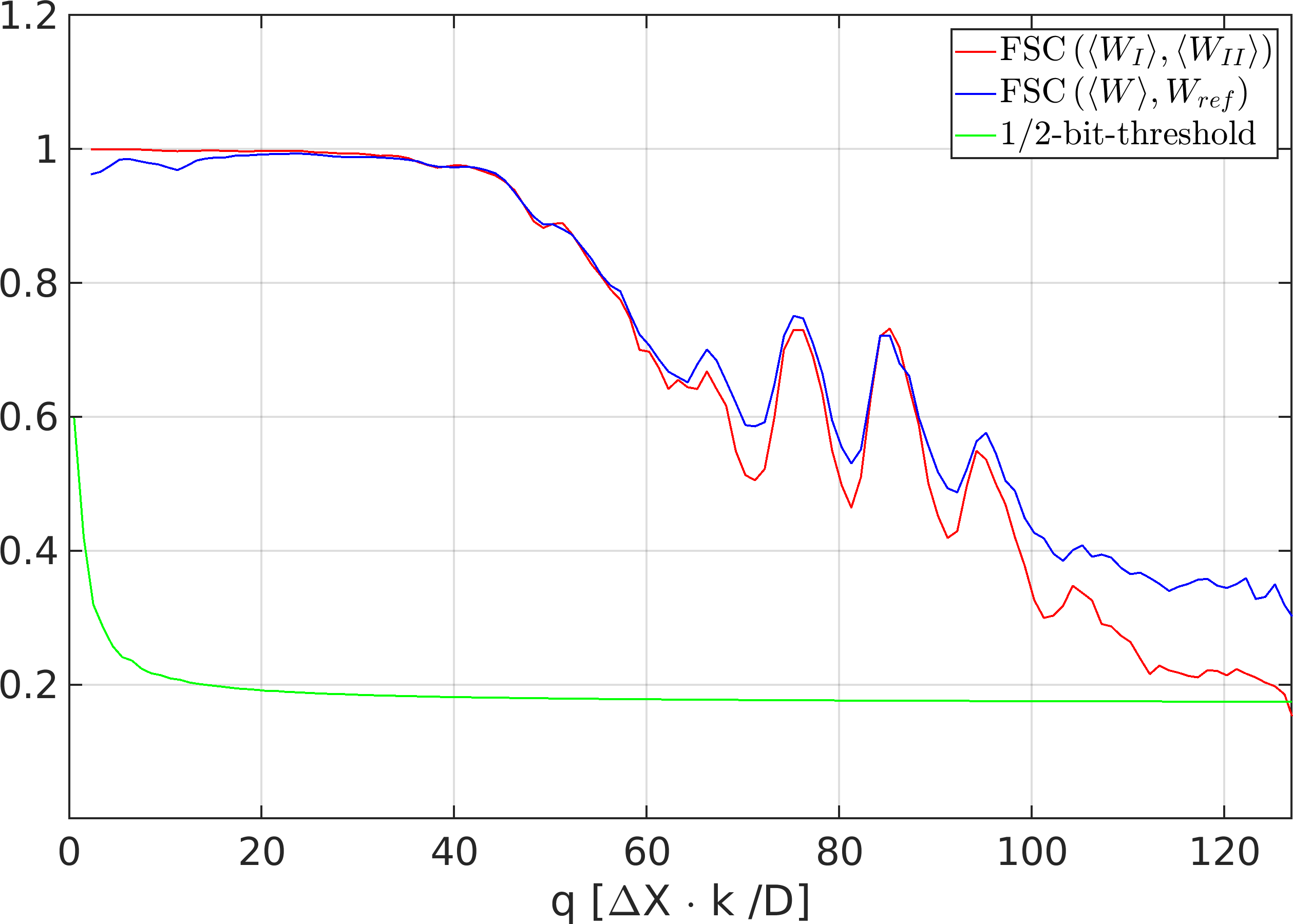}
\caption{The red line illustrates the Fourier shell correlation (FSC) between two EMC-retrieved reciprocal-space volumes resulting from splitting the dataset into two equal halves and performing the same analysis to them as to the whole dataset. The blue line indicates the FSC between $\langle W(q) \rangle$ and $W_\mathrm{ref}$, i.e., the reciprocal-space volume resulting from analyzing the whole dataset using EMC and the reference intensity assembled using known orientations. The green line denotes the half-bit threshold curve, used as a common criterion for resolution determination in analysis of FSC curves.}
\label{fg:fsc} 
\end{figure}

\subsection{Reciprocal space (intensity)}

A comparison of $\langle W(q) \rangle$ with $W_\mathrm{ref}$ is shown in Fig.\ \ref{fg:slices_rec}. Visually, the orthogonal slices through the reconstructed and reference intensities are in very good agreement. This observation is reflected by an overall Pearson correlation coefficient $r = C(W_\mathrm{ref}(\mathbf{q}),\langle W(\mathbf{q})\rangle)$ of $97.1\%$ for $\mathbf{q} \in S$ and $r = 99.0\%$ for $20 \leq q = |\mathbf{q}| \leq 50$ (See the solid lines in Fig.\ \ref{fg:slices_rec}(c) and (f).).

\subsection{Validation (intensity reconstruction)}

To assess the reliability of the reconstructed 3D intensity in reciprocal space we have randomly assigned the frames of the dataset to two independent half-datasets and reconstructed two independent 3D reciprocal space volumes as described before. A Fourier Shell Correlation (FSC) curve \cite{van_heel_fourier_2005}, obtained here directly from the two reciprocal space volumes, is shown in Fig.\ \ref{fg:fsc} (red line). 
It intersects the half-bit threshold curve \cite{van_heel_fourier_2005} at a value beyond $q = 120$, indicating a self-consistent reconstruction of the reciprocal-space volume close to the Nyquist limit. Here, $q$ is measured in units of $\Delta X \cdot k / D$, where $\Delta X = \unit{150}{\micro\metre} $ is the detector pixel pitch, $k$ the wave number and $D$ the sample-detector distance \footnote{Note that in the limit of a flat Ewald sphere patch, which is well-fulfilled here, $q$ then corresponds to the distance to the central beam in units of detector pixels. For further details, see the appendix.}. This analysis  shows the effect of EMC alone rather than merging the effects of orientation determination and phasing, as it would result for doing a traditional FSC analysis on phased results in real space. For comparison, the FSC curve resulting from correlating $\langle W(q) \rangle$ with $W_\mathrm{ref}$, as obtained from known orientations, is also shown (blue line). The high degree of similarity between both curves indicates strong agreement between the result obtained by EMC and the reference intensity distribution, assembled using full knowledge of orientations.

\subsection{Real space (density)}

\begin{figure}
\centering
\includegraphics[width=0.9\columnwidth]{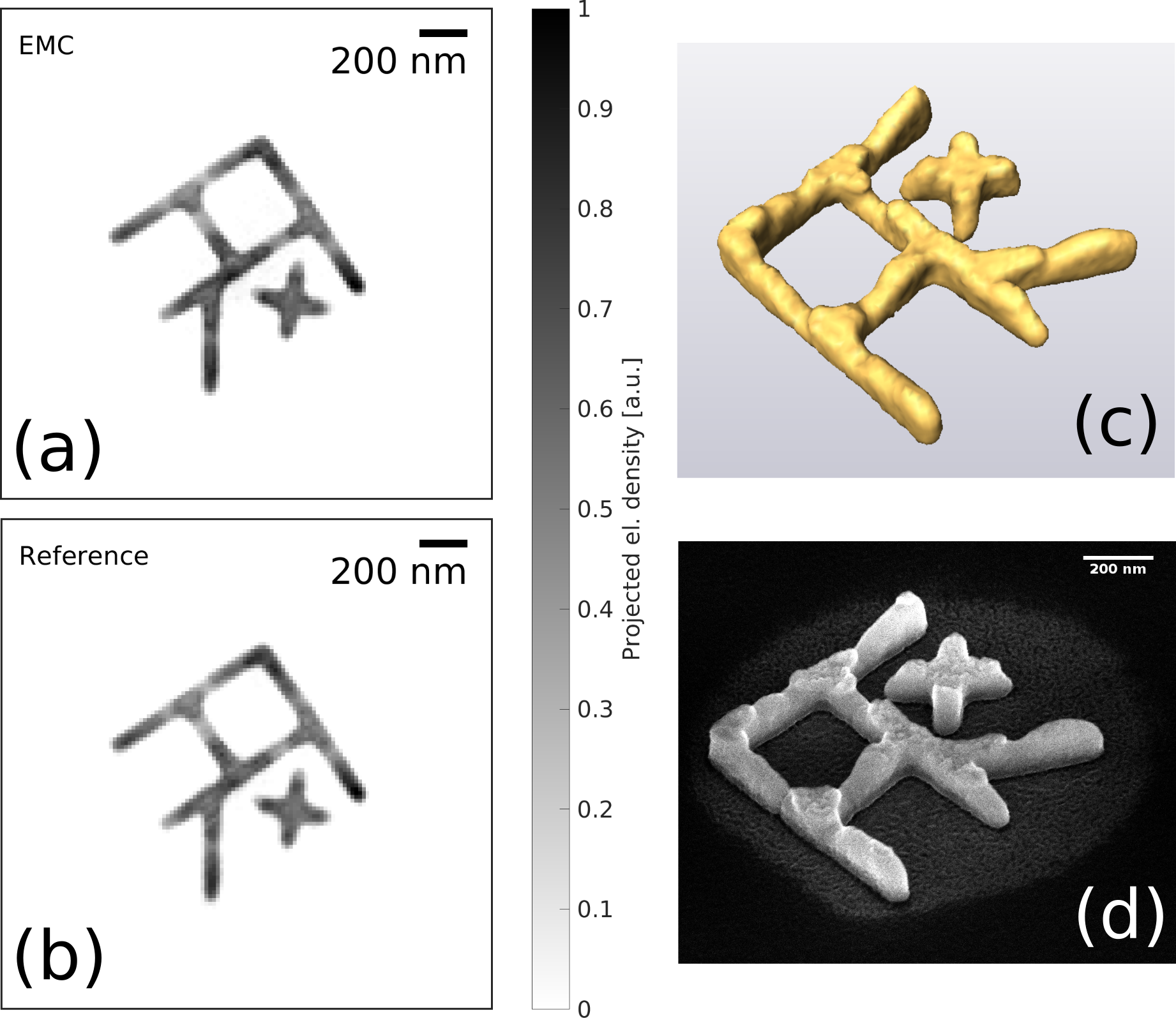}
\caption{Reconstruction of the 3D electron density. (a) Reconstruction from result derived by EMC. The electron density projected along an axis perpendicular to the drawing plane is shown here. (b) Reconstruction from the reference Fourier volume. Again the projected electron density is shown. (c) 3D iso-surface rendering of the reconstructed electron density shown in subfigure (a). The threshold of the iso-surface has been set to $0.2$, given a normalized density with values between 0 and 1. (d) Scanning electron micrograph from the original sample.}
\label{fg:rec}
\end{figure}

To obtain the real-space electron density distribution, the missing phases of the 3D Fourier intensity need to be determined. To this end, we applied standard iterative phase retrieval to the 3D diffraction data, i.e., a combination of the Hybrid-Input-Output (HIO) and the Error Reduction algorithm (ER) \cite{fienup_phase-retrieval_1986,marchesini_x-ray_2003,xiong_gang_coherent_2014}. In total, 600 iterations were applied, i.e.\ 420 iterations of HIO with a feedback parameter $\beta =0.9$, followed by 180 iterations of ER. For further details see the appendix. 

To ensure the reproducibility of the obtained result, 60 reconstructions were performed in total. The results were filtered in a two-step selection process. In a first manual step, we discarded images which visually deviated from the most abundant reconstruction result. In a second step, the remaining  reconstructions (43 for the reference and 37 for EMC data set) were aligned with sub-pixel precision \cite{guizar-sicairos_efficient_2008} and averaged. Then the 20 reconstructions showing the highest correlation with this average were selected for the final average. Note that the procedure applied in the second step could also be used to avoid any manual intervention. However, in such a case, several iterations would likely be required in order to avoid bias by strong outliers in the average. The resulting real-space reconstructions, i.e.\ the real part of the final average, from both reciprocal-space intensities are shown in Fig.\ \ref{fg:rec}.

All details of the reference reconstruction are reproduced in the EMC-based reconstruction down to a resolution level of very few pixels. A comparison of a scanning electron microscopy image of the sample with an iso-surface rendering of the EMC-based reconstruction shows that height variations due to imperfections in the fabrication process are well reproduced by the reconstruction. This identifies the sample as a true 3D structure with features in all coordinate directions.

\subsection{Validation (density reconstruction)}

\begin{figure}
    \centering
    \includegraphics[width=0.95\columnwidth]{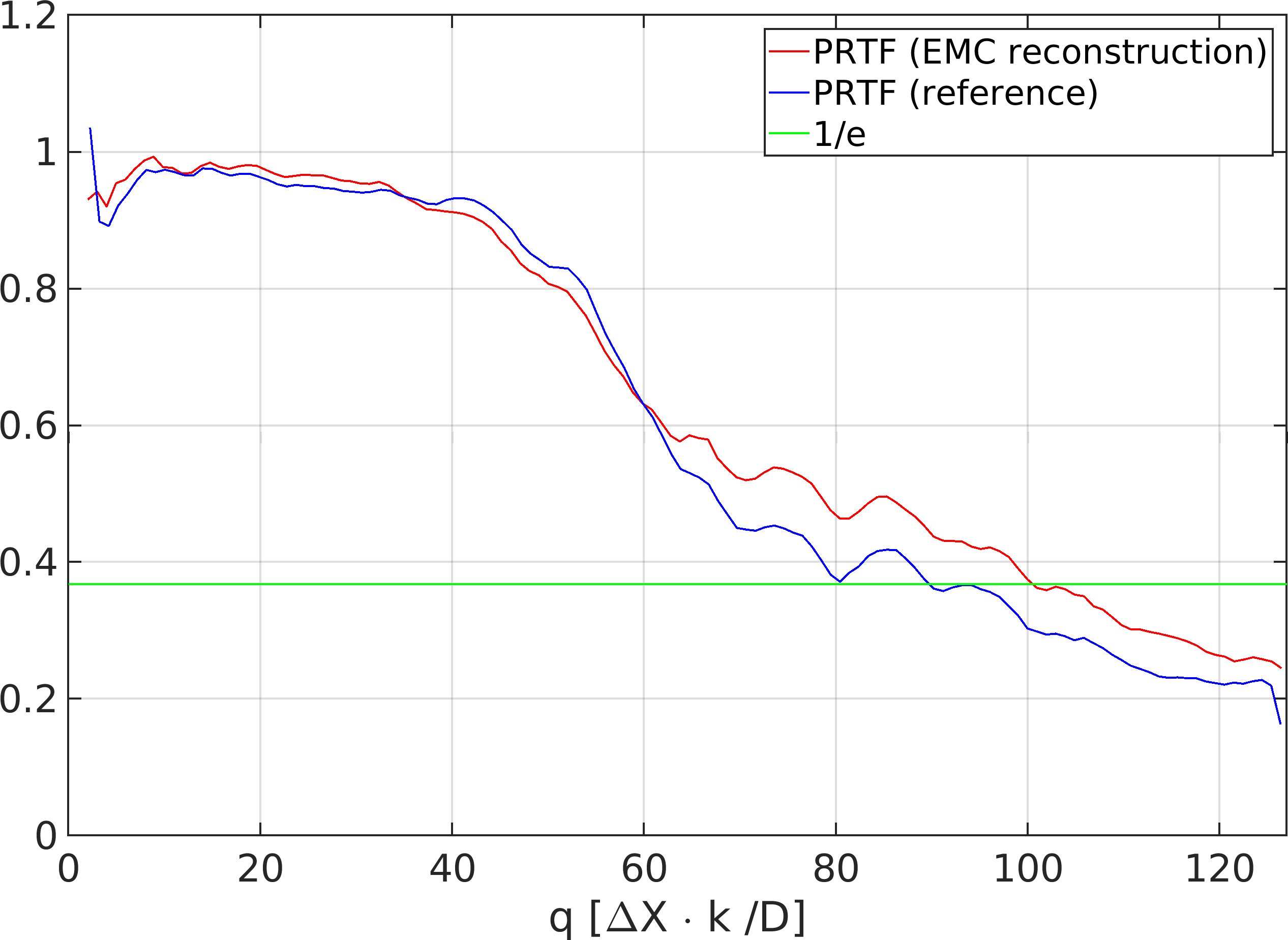}
    \caption{Phase Retrieval Transfer Functions for the reconstructions from the EMC-generated Fourier space intensity ($\langle W(q) \rangle$) and the reference intensity ($W_\mathrm{ref}$). The curves decay to a value of $1/e$ between $q=90$ and $q=100$, corresponding to a half-period resolution between $\unit{20}{\nano\metre}$ and $\unit{23}{\nano\metre}$.}
    \label{fg:prtf}
\end{figure}

The resolution of the final image was estimated via the phase retrieval transfer function (PRTF) according to a procedure similar to the one described in \cite{chapman_high-resolution_2006}. More specifically, before summation of the complex-valued reconstructions, their constant phases were adjusted so that the real part of each reconstruction was maximized. The PRTF curves for the results shown in Fig.\ \ref{fg:rec} are shown in Fig.\ \ref{fg:prtf}. The full-period resolution, as determined by the spatial frequency corresponding to a PRTF-value of $1/e$, amounts to a value between \unit{40}{\nano\metre} and \unit{45}{\nano\metre}. This corresponds to 24 to 26 (full-period) resolution elements within the largest linear extension of the particle, as given by the smallest sphere completely containing the particle (diameter $\approx \unit{1.1}{\micro\metre}$).

\section{Discussion}

\subsection{Significance of the observed level of sparsity}

To assess the significance of the sparsity level for a single detector frame in the present experiment we have calculated the expected average total number of scattered photons outside the central speckle for a selection of $35\,000$ human protein structures from the RCSB Protein Data Bank \footnote{See www.rcsb.org.} (PDB) \cite{berman_protein_2000}. Here a focal-spot diameter of \unit{300}{\nano\metre} was assumed, at a photon energy of \unit{8}{\kilo\electronvolt} and a pulse energy of \unit{1}{\milli\joule} (with 20\% beamline transmission). As a result, it could be shown that under these realistic conditions the minimum diameter for a protein to scatter 50 photons outside the central speckle amounts to \unit{10.6}{\nano\metre}. For further details, see the appendix.

This clearly shows that the signal level in the present experiment, obtained at a synchrotron source from a nano-fabricated gold-structure, is comparable to what can be expected under realistic conditions from a relevant protein structure at an FEL source.

\subsection{Particle complexity, rotation group sampling and Signal-to-Noise ratio}

Another parameter to be discussed is the particle complexity $R$, as measured in half-resolution units per particle radius \cite{loh_reconstruction_2009}. Reconstructing a particle with \unit{10}{\nano\metre} diameter down to a resolution of \unit{3}{\angstrom} results in a complexity of $R\approx 33$, far beyond the current state of the art for FEL-based SPI \cite{kurta_correlations_2017}, i.e.\ $R \approx 7$  for a globular virus particle. The present structure reaches a complexity $>20$ in two dimensions, being constraint in the height direction to a value between 2 and 3. A comparison between the SEM image and an isosurface rendering of the reconstructed particle density shows that all features in the height direction are very well reproduced. Despite its flat shape, this clearly identifies the particle as a true 3D structure and underlines the significance of the present result as a step forward towards the complexity level required for real protein structures.

Furthermore, in comparison to a serial imaging experiment at an FEL, where thousands of particles in random 3D orientations contribute to a full dataset, the number of unique 3D orientations contributing to the present dataset seems relatively low ($M_\mathrm{rot} = 227$). It can be shown, however, that at the given resolution and complexity, this does not restrict the relevance of the result. The required minimum angular separation between adjacent orientations for a sufficient sampling of the 3D rotation group is linked to the complexity $R$ of the particle \cite{loh_cryptotomography:_2010}: $\delta\theta = 1/R$. In case of a non-globular shape, the maximum complexity in a given coordinate direction allows for a conservative estimate, leading here to $\delta \theta = 1/R_\mathrm{max} = \unit{2.2}{\degree}$. This shows that a finer sampling for the tomographic series contributing to the present dataset would not have added more information, at the obtained resolution.

Evidently, the present experiment profits from a high Signal-to-Noise ratio  (see the appendix) which would have been impossible without a setup well-optimized for forward-scattering CDI \cite{chushkin_three-dimensional_2014}. Most importantly, this consists of a set of accurately placed apertures upstream of the sample which are adjusted to define the beam incident on the sample and at the same time to suppress scattering arising from upstream apertures by those further downstream.
Similar schemes can be applied at FEL sources to make them compatible with CDI experiments \cite{munke_coherent_2016}. Even though in the FEL case the aerosol jet in which sample particles are injected through a stream of carrier gas causes an additional source of background scatter \cite{daurer_experimental_2017}, the present dataset gives an experimental benchmark for a signal-to-noise level which likely would allow FEL-based Single Particle Imaging.

\section{Summary and conclusion}

In summary, we have experimentally demonstrated, in the same geometry as used for FEL-based Single Particle Imaging, 
the reconstruction of a complex three-dimensional object using CDI from photon-sparse random projections, at a sparsity level to be expected for a typical protein at an FEL source. To this end, we have collected $454\,000$ data frames with about $\simeq 50$ scattered photons per frame, evenly distributed over $227$ unique orientations, and reconstructed a consistent 3D reciprocal-space volume without explicit knowledge of the orientation of the frames. 

It was shown that, by application of the Expansion-Maximization and Compression (EMC) algorithm, both the reconstructed reciprocal-space intensity and the real-space density of the sample agree to a high level with reconstructions obtained using complete knowledge of frame orientations. We plan to make the dataset freely available in the CXI data bank \cite{maia_coherent_2012}, to be used as a testbed for algorithm development for CDI-based Single Particle Imaging, e.g., by alternative methods for orientation determination. In addition, the dataset can serve as a target for a signal-to-noise level enabling FEL-based SPI in the future.

\appendix
\beginsupplement

\section{Experiment}\label{sc:exp}

Beam-defining slits located approximately $0.5$~m upstream of the sample were set to a gap of \unit{10}{\micro\metre} in both horizontal and vertical directions resulting in a beam size at the sample position of approximately \unit{10}{\micro\metre} $\times$ \unit{10}{\micro\metre}. Diffraction from the beam-defining slits was suppressed by two apertures placed in between the beam-defining slits and the sample \cite{chushkin_three-dimensional_2014}.

Data collection was initiated with a first rotation series spanning a range of $\theta = -80\dots72$ degrees, at one-degree increments. Here, rotations about the coordinate axes by positive angles are defined as left-handed, when looking into the direction of the coordinate axis (see Fig.\ 1 in the main text). Orientations between $\theta = 46$ and $\theta=48$ were unintentionally omitted during the process of data colletion which could only be semi-automated to ensure continuous centering of the sample in the beam. Some of the frames collected at $\theta = -16$ were not saved correctly. Therefore, the data for this orientation was excluded from the analysis, so that data from 149 unique orientations from the first rotation series were used for analysis.

After completion of the first rotation series, the sample was manually removed from the sample stage, rotated by $\chi = -99.37$ degrees about the $z$-axis, and returned to the sample stage. The latter angle was determined \textit{a posteriori} from the correlation between summed diffraction patterns at $\theta = 0$. The second rotation series consisted of $78$ orientations, spanning a range of $\theta = -82\dots72$ degrees, with 2-degree increments. 

The ESRF synchrotron was operated in four-bunch-mode, with each bunch carrying around $10$~mA of maximum current. With a resulting bunch frequency of $1.42$~MHz and with the given attenuation less than 100 photons reached the detector from a single bunch. For a pure counting detector the photons from a single bunch arrive in a time span far too short to be discriminated by the counting electronics. The MM-PAD uses a charge integrating front-end with an extended dynamic range that is achieved by removing a known charge from the pixel input node when the integrator output nears saturation. The number of charge removals is tracked with an in-pixel 18-bit counter. Each charge removal is approximately equivalent to $200$ photons at $8.1$ keV. The MM-PAD can accommodate an instantaneous x-ray pulse up to this level of 200 photons per pixel per bunch without saturation, allowing, in this experiment, receiving the direct beam on the detector without a central stop. This allowed precise optimization of the beamline settings to suppress parasitic slit scattering and gives the user a large flexibility in selecting software masks to exclude certain detector regions from the subsequent analysis steps, such as those dominated by the central beam.

\section{Data analysis}

\subsection{Detector calibration}

The raw signal output for each pixel from the MM-PAD is given in analog-to-digital units (ADUs) which are proportional to the number of electron-hole pairs produced in the Si sensor material of the detector. As further data analysis here requires calibrated detector data, i.e., the number of photons per pixel per frame, the raw signal first needs to be converted accordingly. 

Before calibration, a small number of malfunctioning pixels were identified based on their noise level at zero-photon illumination: The root-mean-square (RMS) noise level in each pixel was determined from 15200 dark exposures, equally distributed over the time it took to collect the data analyzed here. Each dark frame was taken with the same exposure settings as the diffraction data frames. A pixel was identified as malfunctioning (and masked out), if its RMS value deviated by more than 10\% from the mean RMS value, averaged over all pixels and all patterns. Also, the detector consisted of 6 detector tiles in a $2\times3$ arrangement. There were gaps between tiles that were insensitive to x-rays and were, therefore, masked out. In all, $6.7\%$ of 105,336 pixels in the whole detector area were masked out for further analysis, including two pixels closest to the beam.

The subsequent calibration process can be described as follows. As a first step, a dark frame was subtracted from each measurement frame. Since dark frames are measurements of the detector output in the absence of signal, and thus subject to the same read noise as other measurements, a common procedure to reduce the noise associated with dark frame subtraction is to define dark frames as an average of many frames, in this case 200. Dark frames are usually static over the short term. Longer term drift is accounted for by periodically updating the subtracted dark frame with new measurements.
    
Secondly, the gain was determined. To this end, 2000 representative frames from a measurement were used to generate a histogram of raw count rates. As the vast majority of pixels received zero photons during a measurement, a region of interest was defined by selecting those pixels with an average count rate between 5 and 15 ADUs. This defines an interval which is roughly centered around the expected number of ADUs for a single $8.1$-keV photon \cite{giewekemeyer_high-dynamic-range_2014}.

A Gaussian fit to the left-most (zero-photon, or noise peak) in the histogram resulted in a noise level (standard deviation) of $2.1$~ADU. A small offset $<1.0$ ADU in the position of the noise peak was determined here as well and corrected for. Subsequently, a Fourier analysis was applied to the histogram to determine the peak separation, yielding a gain of $11.1$ ADU for a single $8.1$-keV photon. This implies a signal-to-noise (SNR) value of $5.2$ at $\unit{8.1}{\kilo\electronvolt}$. 

As described previously \cite{ayyer_real-space_2014}, a threshold $E_t$ can be applied to discriminate single photon events from noise. This step is the central procedure of the detector calibration and is especially relevant, if the data are very sparse, as in the present case. Applying such a threshold inevitably leads to a certain amount of false events, i.e., the detection of a photon where there was none, and vice versa. A natural choice for the threshold parameter is given by the condition that $P(1|0) = P(0|1)$, i.e., the probability of detecting at least one photon when there is none (false positives) equals the probability of detecting no photons, if there is at least one (false negatives). Neglecting multiple-photon events and assuming a Gaussian noise distribution, this condition is given for $E_t = E_\gamma/2$. Note, however, that in this case $P(1|0) = 1/2 \textrm{erfc}(E_t/\sqrt{2}\sigma) \simeq 4.7\cdot10^{-3}$ \cite{becker_single_2012}. For the present case, with around $40,000$ active detector pixels in the region of interest of a single frame (see below), this would already lead to a false-positive rate of around $180$ events per frame. This is far too high, if the expected signal is on the order of $50$ photons per frame. A previous study using EMC for reconstruction of real-space tomographic data used $E_t = 0.6E_\gamma$ \cite{ayyer_real-space_2014}. We have decided here to use $E_t = E_\gamma - \textrm{HWHM}_n$ where $\textrm{HWHM}_n = \sqrt{2\textrm{ln}(2)}\sigma$ is the Half Width at Half Maximum of the noise peak. In numbers, this leads to $E_t \simeq 0.77E_\gamma$, so that $P(1|0) = 3\cdot10^{-5}$. As a consequence, the expected number of false-positive events is lowered to less than 2 events per frame. Note that with the chosen threshold we tolerate a false-negative probability $P(0|1) \simeq 0.12$.

As the photon distribution in the data frames is very sparse, cosmic rays are often strongly visible against the low background, especially at high diffraction angles. They usually lead to characteristic streaks several pixels long, with count rates equivalent to a few 8.1-keV photons per pixel. Even though a large body of methods do exist for removing cosmic rays \cite{farage_evaluation_2005}, here we utilized the sparse nature of the data to remove them in a simple statistical manner. 
    
Assuming Poisson statistics, the probability for a given pixel $i$ with expectation value $\lambda_i$ to receive more than a single photon, is given by $P_{\lambda_i}(X \geq 2) = 1 - P_{\lambda_i}(X < 2) = 1-\exp(-\lambda_i)(\lambda_i+1)$. To discriminate counts due to cosmic rays from sample diffraction we apply a twofold test on each pixel. First, we determine the expectation value $\lambda_i$ for pixel $i$ from its mean over all frames. Secondly, if $\lambda_i \leq \lambda_{\rm th}$ for a threshold expectation value $\lambda_{\rm th}$, we regard any count value $X>1$ as originating from a cosmic ray and set its value to $0$.  If $\lambda_i > \lambda_{\rm th}$, we leave the pixel unchanged. The choice of $\lambda_{\rm th}$ determines the maximum error that we will make during this process, i.e., the maximum number of false identifications per pixel per frame. This includes false deletions of values which originate from sample diffraction (false negatives) and vice versa (false positives). If we accept at maximum one false deletion of a pixel's value per orientation, i.e., per 2000 frames, we can set $P_{\lambda_i = \lambda_{\rm th}}(X \geq 2) = 1/2000$. As here $\lambda_{\rm th} << 1$, we can approximate $P_{\lambda_{\rm th}}(X \geq 2) \approx 1-(1-\lambda_{\rm th})(1+\lambda_{\rm th}) = \lambda_{\rm th}^2$ and therefore $\lambda_{\rm th} \approx 1/\sqrt{2000} \approx 0.0224$. 
    
Note that for the majority of pixels $\lambda_i$ is much smaller than $\lambda_{\rm th}$, so that the average number of false positive cosmic ray identifications per orientation is much lower than 1. To determine the average number of false negatives per orientation is more difficult, as this number depends on the expectation value of counts due to cosmic rays, per pixel per frame. In general, the influence of cosmic rays becomes much less relevant as $\lambda_{\rm th}$ grows. Therefore, we regard their contribution to the total count rate as negligible in this area \footnote{Another discrimination could be based on a comparison of $\lambda_i^{\rm (diffraction)}$ with $\lambda_i^{\rm (cosmic)}$ for each pixel.}.
            
\subsection{Mapping pixel coordinates to Ewald sphere coordinates}

For a detector (field of view) with a width of $N_x$ pixels and a height of $N_y$ pixels, distributed here on a Cartesian grid, each pixel is represented by a linear index $i = 1,\dots,N_x \times N_y$ which may be mapped to two-dimensional indices $(n_y,n_x)$ according to
\begin{align}
	n_y(i) &= \lceil i/N_y\rceil -L-1&\quad\text{(row index)}&\\
    n_x(i) &= (i-1) \,\text{mod}\, N_x -L&\quad\text{(column index)}&.
\end{align}
Here, $N_x = N_y = 2L+1$ with $L$ denoting the distance from the central pixel to the pixel at the edge of the field of view (in pixel units). With this definition, $n_{x,y} \in \{-L,\dots,L\}$.

In each pixel $(n_y,n_x)$ the diffraction signal at a certain location on the Ewald sphere is measured, whose reciprocal space coordinate is given by Ref.~\onlinecite{loh_reconstruction_2009}
\begin{align}
    \label{eq:es1}
   	q_{(x,y)} &= \frac{n_{(x,y)}}{\sqrt{1+\left(n_x^2+n_y^2\right)/(L\cot\Theta)^2}}\\ 
   	q_z &= \frac{L\cdot\cot\Theta}{\sqrt{1+\left(n_x^2+n_y^2\right)/(L\cot\Theta)^2} } - L\cdot\cot\Theta,
   	\label{eq:es2}
\end{align}
with $\Theta = \arctan (L\cdot \Delta X / D)$. Here $\Delta X$ denotes the pixel pitch of the detector and $D$ the distance of the sample to the detector plane. In this description, the unitless reciprocal space coordinates $(q_x,q_y,q_z)$ are related to their unit-carrying counterparts $q_i'$ via $q_i = q_i' \cdot D/(k\cdot \Delta X )$. Here $k$ denotes the wave number $k = 2\pi/\lambda$ with photon wavelength $\lambda$. For simplicity, the coordinate index $i$ will henceforth be omitted.

\subsection{Discretization of Fourier and real space}

For compatibility with standard implementations of the Discrete Fourier Transform (DFT) a 3D Cartesian grid is defined 
in Fourier space with cubic voxels of unit sidelength in dimensionless units, i.e., $\Delta q = \Delta q' \cdot D/(k\cdot \Delta X) = 1$. The grid is defined within a cube of sidelength $M = 2q_\mathrm{max}+1$. A maximum value of  
\begin{equation}
    q_\mathrm{max} = \left\lceil 2D/\Delta X \cdot \sin\left(\frac{1}{2}\arctan\left(\frac{L\cdot \Delta X}{D}\right)\right) \right\rceil
\end{equation}
then corresponds to diffraction to the edge of the (region of interest on the) detector. Using this value, it is assured that a circular patch of the Ewald sphere, with a radius corresponding to the distance from the center to the edge of the detector, is always contained within the gridded cube mentioned above. For the present experiment we have chosen $L$ such that $q_\mathrm{max} = 127$.

The reciprocity relation of the DFT, $\Delta q' = 2\pi / (M \Delta x')$, gives access to the grid spacing of the corresponding gridded cube in real space, namely $\Delta x' = \lambda D / (M \Delta X)$. Similar to Fourier space, a unitless grid can be defined with lengths measured in units $\Delta x'$ which correspond to DFT-based resolution units defined by the maximum q-vector that is reached by an edge pixel of the detector. Note that these resolution units set a lower (best) limit to the physical resolution of the experiment, but are not necessarily equal to the latter. The dimensionless particle radius $R$ is related to the physical particle radius $R'$ via $R' = R\cdot a$. $a = \Delta x'$ is used in Ref.~\onlinecite{ayyer_dragonfly:_2016}. For easier comparison to experimental data, we choose $a = \Delta r$ where $\Delta r$ corresponds to a half-period resolution element, as obtained from an analysis of reconstructed resolution (see below).

\subsection{Definition of binary detector masks}

\begin{figure}
\centering
\includegraphics[width=0.9\columnwidth]{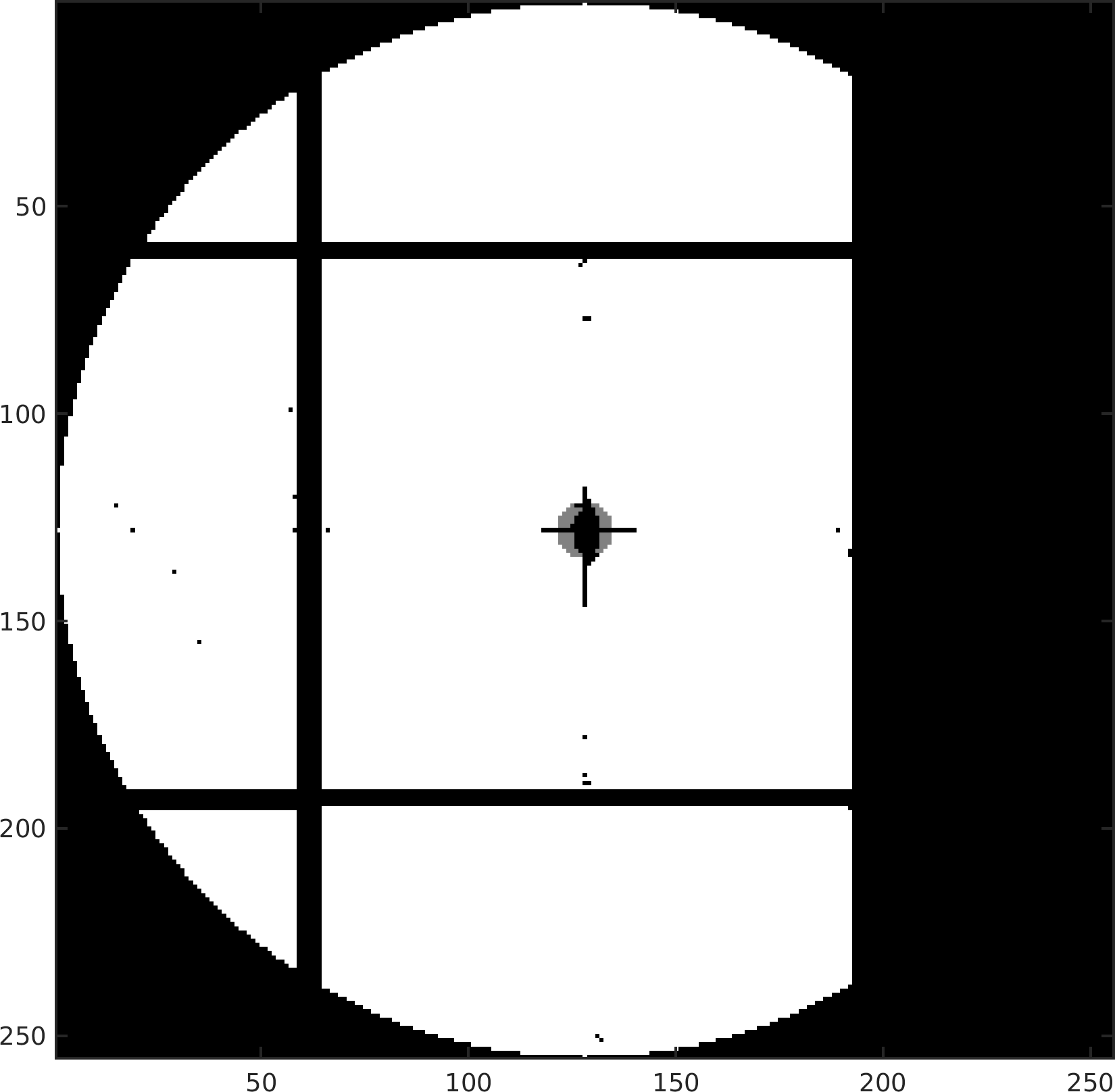}
\caption{Detector mask sized 255 by 255 pixels, showing those pixels in white and gray which were used from each frame as an input to EMC. Grey pixels near the center where excluded from orientation determination in the maximization step in EMC, as they still contain mostly non-scattered photons from the primary beam. Axes labels denote pixel numbers.}
\label{fg:masks}
\end{figure}

To optimize the orientation determination using EMC, several binary masks have been defined which describe certain properties for each pixel. 
    
Given a set $M_0 = \{i  = 1,\dots,N_x\times N_y\}$ of pixels with linear index $i$, the most basic mask of valid pixels is given by the set $M_v \subset M_0$ of all pixels which do not fall on non-sensitive areas between detector modules and are considered as working properly as defined above. 

Secondly, a `beam mask' $M_b \subset M_v$ was defined that excludes all pixels whose signal is dominated by the non-scattered beam or background radiation. More precisely, 
\begin{equation}
    M_b = 
        M_v \setminus \left\{i \in M_v \mid \lambda_i^{(\textrm{sa})} / \lambda_i^{(\textrm{bg})} < r \wedge \lambda_i^{(\textrm{bg})}\cdot N_{\textrm bg} > 25 \right\}
\end{equation}
Here $\lambda_i^{(\textrm{bg, sa})}$ is the mean value of pixel $i$ for background (empty-beam) and sample data, respectively and $N_{\textrm bg}$ is the number of frames to define the mean of the background. Consequently, a pixel is considered as dominated by background (including the non-scattered beam), if its signal-to-background value is smaller than $r$ and the mean value of the background itself has been determined with an SNR of at least 5 (Rose-criterion), assuming Poisson statistics. Here we choose $r=2$ and $N_{\textrm bg}$ has a value of $450000$ in the present case. 

Thirdly, for application of EMC, a mask $M_{EC} \subset M_b$ was defined that includes those pixels to be considered within the expansion and the maximization step of EMC. This mask is defined with respect to the reciprocal space coordinate values $(q_x(i),q_y(i),q_z(i))$ of each pixel on the Ewald sphere, to include only those pixels which correspond to a spherical cap. More precisely,
\begin{align}
   	M_{EC} = M_b \setminus \left\{i \in M_b \mid q_x(i)^2+q_y(i)^2+q_z(i)^2 > q_\mathrm{max}^2 \right \}
\end{align}
with 
\begin{equation}
    q_\mathrm{max} = \left\lceil\max_i\left\{ \sqrt{q_x(i)^2+q_z(i)^2},\sqrt{q_y(i)^2+q_z(i)^2}\right\}\right\rceil.
\end{equation}

Lastly, a mask $M_{M} \subset M_{EC}$ of pixels was created which defines the pixels which contribute to the 3D diffraction volume but which are not considered for orientation determination within the maximization step \cite{ayyer_dragonfly:_2016}. This allows, for example, to exclude pixels near the center which still have a significant amount of signal from the direct beam which could have a detrimental effect on orientation determination. More specifically, $M_M$ was defined here as
\begin{equation}
	M_M = M_{EC} \setminus \left\{i \in M_{EC} \mid q_x(i)^2+q_y(i)^2+q_z(i)^2 < q_\text{min}^2\right\}.
\end{equation}
with $q_\text{min} = 7$.

The mask $M_M$ is illustrated in Fig.\ \ref{fg:masks} by white pixels, whereas the difference set $M_{EC}\setminus M_M$ is indicated by gray pixels.

\subsection{Quaternions and rotation series}

Following \cite{loh_cryptotomography:_2010}, let $W(\mathbf{q})$ denote the integrated scattering intensity at reciprocal space coordinate $\mathbf{q}$, with the particle fixed in a unique reference orientation. For the given geometry, each detector pixel with index $i$ denotes a unique sampling point $\mathbf{q} = \mathbf{q}_i$ on the Ewald sphere, as described by Eqs.\ (\ref{eq:es1}) and (\ref{eq:es2}). With the sample in the original reference orientation, $W(\mathbf{q}_i)$ then samples one point of the reciprocal space intensity distribution $W$. 

The goal of the experiment is to homogeneously sample $W$ by changing the relative orientation of the sample with respect to the Ewald sphere. According to Euler's rotation theorem, any orientation of the sample with respect to a given reference orientation can be described by a rotation by angle $\phi$ about a single axis $\mathbf{n}$ \footnote{For $\alpha>0$ we here define rotations to be counter-clockwise, looking into the direction of $\mathbf{n}$.}. Taking the perspective of a fixed sample and a rotating Ewald sphere, the transformation of sampling point $\mathbf{q}_i$ on the Ewald sphere then has to follow the inverse rotation. Thus, if $R_{\mathbf{n}}(\phi)$ denotes the rotation matrix that describes the rotation of the sample (coordinates), then $R_{\mathbf{n}}(\phi)^{-1} = R_{\mathbf{n}}(-\phi)$ describes the corresponding rotation of Ewald sphere coordinates (for a fixed sample). I.e., the new sampling point is given as $W(R_{\mathbf{n}}(-\phi)\mathbf{q}_i)$, or, more generally, $W(R^{(ES)}_j\mathbf{q}_i)$ for orientation $j$.

EMC implements the determination of a the sample's orientation for a given data frame as a statistical search within a given list of orientations \cite{loh_reconstruction_2009}. This list is one of the inputs for the algorithm. For the general case of a freely rotating particle, the possible orientations have to be uniformly spread in the space of 3D rotations (the 3D rotation group $SO(3)$). In the present experiment the free 3D rotation is replaced by two tomographic series (see Section~\ref{sc:exp}), with two different, nearly orthogonal rotation axes, so that in total an almost complete coverage of diffraction space is achieved. 

Orientation $j$ of the sample during the first series can be described by a rotation matrix $R^{(S)}_j = R_z(\theta_j)$. Here $\theta_j$ corresponds to the angles as defined in Section \ref{sc:exp}. The corresponding rotations of the Ewald sphere are thus described by
\begin{equation}
    R^{(ES)}_j = [R^{(S)}_j]^{-1} = R_z(-\theta_j).
    \label{eq:rs1}
\end{equation}
The second series involves sample rotations about two axes (first about $z$-, then about $y$-axis), its rotation matrices are given by $R^{(S)}_j = R_y(\theta_j)R_z(\chi)$. The corresponding transformation of the Ewald sphere coordinates is then represented by the matrix
\begin{equation}
    R^{(ES)}_j = [R^{(S)}_j]^{-1} = R_z^{-1}(\chi)R_y^{-1}(\theta_j) = R_z(-\chi)R_y(-\theta_j).
    \label{eq:rs2}
\end{equation}
EMC uses the quaternion formalism to describe 3D rotations \cite{loh_reconstruction_2009}. The quaternion that encodes the rotation by an angle $\phi$ about an axis $\mathbf{n}$, with respect to a fixed reference orientation, is given by 
\begin{align}
	\mk{u}(\phi,\mathbf{n}) = (\cos(\phi/2),\sin(\phi/2) \mathbf{n}).
	\label{eq:aa_rep}
\end{align}
Here $\mk{u}_0 = \cos(\phi/2)$ denotes the `scalar' part and $\mathbf{u} = \sin(\phi/2) \mathbf{n}$ the `vector' part of the quaternion. A quaternion norm $\lVert \mk{u} \rVert = \sqrt{\mk{u}\cdot \mk{u}}$ can be defined using the standard scalar product $\mk{u}\cdot \mk{v} = \mk{u}_0\mk{v}_0 + \mathbf{u}\cdot\mathbf{v}$. For counter-clockwise rotations defined as positive, a rotated vector $\mathbf{r}' = \mathbf{R}\mathbf{r}$ may then be obtained by 
\begin{align}
	\mk{r}' = \mk{u}^{-1}\,\mk{r}\,\mk{u} =: \mathbf{R}_{\mk{u}}(\mk{r})
	\label{eq:quat_rot}
\end{align}
with $\mk{r} = (\mk{r}_0,\mathbf{r})$ and $\mk{r}_0 = \mk{r}'_0 = 0$. Here,  quaternion multiplication has to be used on the right-hand side of the equation, i.e., $\mk{u}\,\mk{v} = (\mk{u}_0\mk{v}_0-\mathbf{u}\cdot\mathbf{v},\mk{u}_0\mathbf{v}+\mk{v}_0\mathbf{u}+\mathbf{u}\times\mathbf{v})$. Upon application of Eq.\ (\ref{eq:quat_rot}), one may obtain the rotation matrix $\mathbf{R}(\mk{u})$ using elements of $\mk{u}$ (see Ref.~\onlinecite{loh_reconstruction_2009}, Eq.\ (C1)). Sequential application of Eq.\ (\ref{eq:quat_rot}) further implies that the application of two rotations in the order $1 \rightarrow 2$, can be described by the quaternion product $\mk{u}_1\mk{u}_2$ (in this order):
\begin{equation}
    \mathbf{R}_{\mk{u}_2}(\mathbf{R}_{\mk{u}_1}(\mk{r})) = \mathbf{R}_{\mk{u}_1\mk{u}_2}(\mk{r}).
\end{equation}
For the corresponding rotation matrices that act on the vector part of $\mk{r}$ only, we thus have
\begin{equation}
    R(\mk{u}_2)R(\mk{u}_1) = R(\mk{u}_1\mk{u}_2).
    \label{eq:qm}
\end{equation}

The first rotation series applied in the present experiment may be described by quaternions $\mk{u}_j(-\theta_j,\mathbf{e}_y)$ (see Eq. (\ref{eq:rs1})). The quaternions $\mk{p}_j$ for the second series may be obtained as a combination of a constant quaternion $\mk{v}$ that describes the new sample orientation at the start of the rotation series, and another quaternion $\mk{u}_j$ that changes for every orientation in the series. With Eqs.\ (\ref{eq:rs2}) and (\ref{eq:qm}):
\begin{eqnarray}
    R^{(ES)}_j &=& R_z(-\chi)R_y(-\theta_j) \nonumber\\
               &=& R(\mk{v}(-\chi,\mathbf{e}_z)R(\mk{u}(-\theta_j,\mathbf{e}_y))\nonumber\\
               &=& R(\mk{u}(-\theta_j,\mathbf{e}_y)\mk{v}(-\chi,\mathbf{e}_z)).
\end{eqnarray}
Thus, $\mk{p}_j = \mk{u}(-\theta_j,\mathbf{e}_y)\mk{v}(-\chi,\mathbf{e}_z)$. To first approximation (from experimental parameters), $\mk{v}(-\chi,\mathbf{e}_z) = (\cos(-\chi/2), 0 ,0, \sin(-\chi/2))$, with $\chi = -99.37$.

\subsection{Alignment of rotation series} \label{sc:alignment}

If the orientations are known, all $W_{ij} = W(\mathbf{R_j}\mathbf{q_i})$ may be interpolated to a 3D diffraction volume $W(\mathbf{p})$ on an equi-spaced Cartesian grid $\mathbf{p}$ in Fourier space, as applied in the compression step of EMC \cite{loh_reconstruction_2009}. However, it was observed here that the Fourier space intensities $W_1(\mathbf{p})$ and $W_2(\mathbf{p})$, resulting from the two rotation series do not ideally match, using $\mk{v}$ as defined from the nominal experimental values: By removing the sample frame from its holder, turning it manually about an axis perpendicular to its surface and then returning it to its holder, it is likely that an orientation change slightly different from the expected one has been performed: In the nominal zero-position of the two rotation series the sample frame was likely not exactly perpendicular to the optical axis.

Therefore, we have applied an iterative optimization of $\mk{v}$, based on maximizing the two-point correlation $C_{12}$ between $W_1(\mathbf{p})$ and $W_2(\mathbf{p},\mk{v})$. Here the argument $\mk{v}$ indicates the dependence of $W_2$ on $\mk{v}$. This process is equivalent to orientational registration of two scalar fields in three dimensions. For each iteration $n$, we generated a set of $N$ candidate quaternions 
\begin{align}
	V_n = \left\{\mk{v} \mid d(\mk{v},\mk{v}_{n-1}) < r_n\right\}
\end{align}
with $\mk{v}_{n-1}$ denoting the best estimate for iteration $(n-1)$. $d(\mk{u},\mk{v})$ for two unit quaternions $\mk{u}$ and $\mk{v}$ is defined here as
\begin{align}
	d(\mk{u},\mk{v}) = \min\{2\arccos(\mk{u}\cdot \mk{v}),2\arccos(-\mk{u}\cdot \mk{v})\},
\end{align}
representing a metric on $SO(3)$ \cite{lavalle_planning_2006, huynh_metrics_2009} and a pseudometric on $S(3)$, the space of unit quaternions. Note that for $\arccos(\mk{u}\cdot\mk{v}) < \pi$, $d(\mk{u},\mk{v}) \in [0,\pi]$ is equal to the rotation angle $\phi$ of the quaternion that maps $\mk{u}$ onto $\mk{v}$ and vice versa. In order to obtain an unbiased set of random quaternions a standard algorithm for uniform sampling of $SO(3)$ was applied \cite{kuffner_effective_2004}. In order to resolve the 2-to-1 mapping from S(3) to SO(3), $\mk{u}$ was replaced with $-u$, if $d(\mk{u},\mk{v}) <r_n$ for a given $\mk{u}$ and $\arccos(-\mk{u}\cdot \mk{v}) < \arccos(\mk{u}\cdot \mk{v})$. Note that $\mk{u}$ and $-\mk{u}$ define the same rotation, as can be seen by application of Eq.\ (\ref{eq:quat_rot}).

For each iteration, then the quaternion $\mk{v}_n \in V_n$ was determined which maximizes $C_{12}$
and $r_{n+1}$ was set to $r_{n+1} = d(\mk{v}_n,\mk{v}_{n-1})$. The algorithm was started with $\mk{v}_0 = \mk{v}$ as defined above and $r_1 = \pi/180 \cdot 20$. The algorithm was stopped when $r_n \leq \pi/180 \cdot 1$. To accelerate the calculation of $C_{12}$ only voxels within a shell defined by a radius $R_{\min}=10 \leq R \leq R_{\max}=60$ were considered. To speed up the calculation of $C_{12}$ for all candidate quaternions $\mk{v}$, $W_2$ was not built up from linear interpolation of all Ewald sphere slices for each $\mk{v}$, but it was formed via building up $W_2$ once for $\mk{v} = 0$ and subsequent rotation of the corresponding Fourier space distribution by a matrix corresponding to the current candidate $\mk{v}$. The latter step was performed using the Matlab routine \texttt{imwarp.m}, being part of the Image Processing Toolbox \footnote{Matlab R2017b, The MathWorks, Nattick (MA), USA (2017).}. 

For the optimized $\mk{v}_{\textrm{opt}}$ we obtained $d(\mk{v}_{\textrm{opt}},\mk{v}) \approx 11$ deg with respect to the nominal $\mk{v} = (\cos(-\chi/2),0,0,\sin(-\chi/2))$, and the rotation axes of $\mk{v}_{\textrm{opt}}$ and $\mk{v}$ differ by approximately $\arccos(\mathbf{v}_{\textrm{opt}}\cdot{\mathbf{v}}/(\sin(\arccos(\mathit{v}_0))\sin(\arccos({\mathit{v}_{\textrm{opt}}}_0)))\simeq 7$ deg. After registration, no visual inconsistencies between $W_1$ and $W_2$ were observed any more (see Fig.\ 3 (d-f) of the main text).

\subsection{Friedel symmetrization in EMC}

EMC includes an optional Friedel symmetrization step after each iteration, enforcing $W(\mathbf{q}) = W(\mathbf{-q})$ for the current iterate of the 3D Fourier space model $W(\mathbf{q})$ \cite{loh_reconstruction_2009}. This is often justified due to negligible absorption within the sample. The 3D support mask $S(\mathbf{q})$ identifies all voxels in Fourier space which are reached by a measurement, i.e., an Ewald sphere slice. The set of orientations used in the present experiment defines a support which is not centrosymmetric. Therefore, the standard Friedel symmetrization step in EMC was adapted to the following procedure:
\begin{equation}
    W'(\mathbf{q}) = 
    \begin{cases}
        \left(W(\mathbf{q}) + W(-\mathbf{q})\right)/2 & \textrm{for } \mathbf{q} \in S \wedge \mathbf{-q} \in S \\
        W(\mathbf{q}) & \textrm{for } \mathbf{q} \in S \wedge \mathbf{-q} \not\in S.
    \end{cases}
\end{equation}

\section{Results}

\subsection{Signal versus Background}

\begin{figure}
\includegraphics[width=0.9\columnwidth]{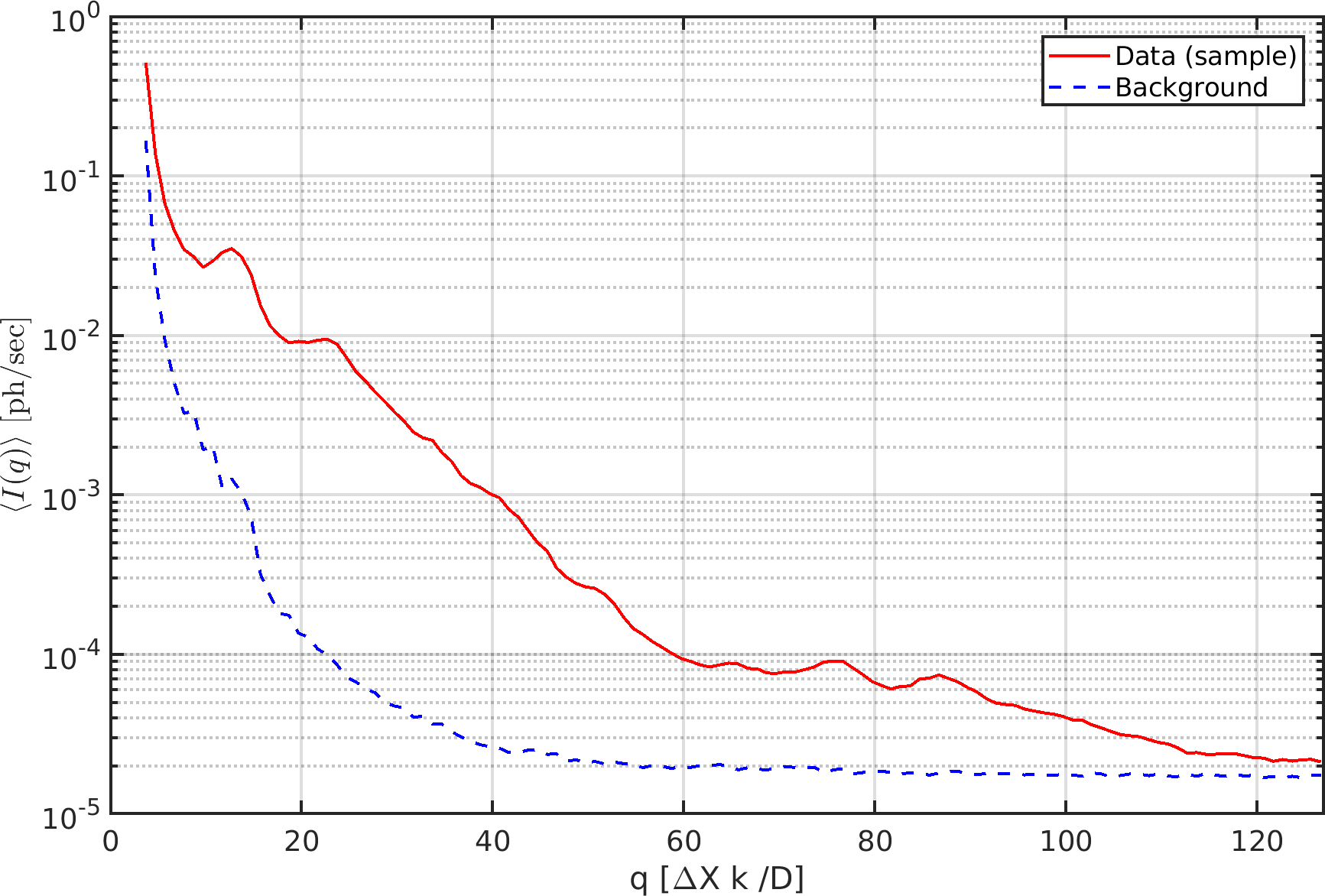}
\caption{Azimuthally averaged mean signal from the sample and background versus the dimensionless radial coordinate $q$ (see above and main text).}
\label{fg:sbr}
\end{figure}

To assess the influence of instrumental background on the data, the azimuthally averaged scattering signal from the sample (including other sources of scatter) and the instrumental background is plotted in Fig.\ \ref{fg:sbr}. For averaging, the mean signal $\lambda_i^\mathrm{(bg,sa)}$, as introduced above, has been used. Quite generally, the signal from the sample dominates the background, with a difference in magnitude from about two orders of magnitude at medium-range $q$ down to much less than one order of magnitude at very small and very high $q$. Here as well as for the calculation of the Fourier Ring correlation plots (see main text) we have used histogram-based azimuthal averaging using the same principle as described in Ref.~\onlinecite{kieffer_pyfai:_2014}.

\subsection{Analysis of retrieved Fourier space results}

\begin{figure}
    \centering
    \includegraphics[width=\columnwidth]{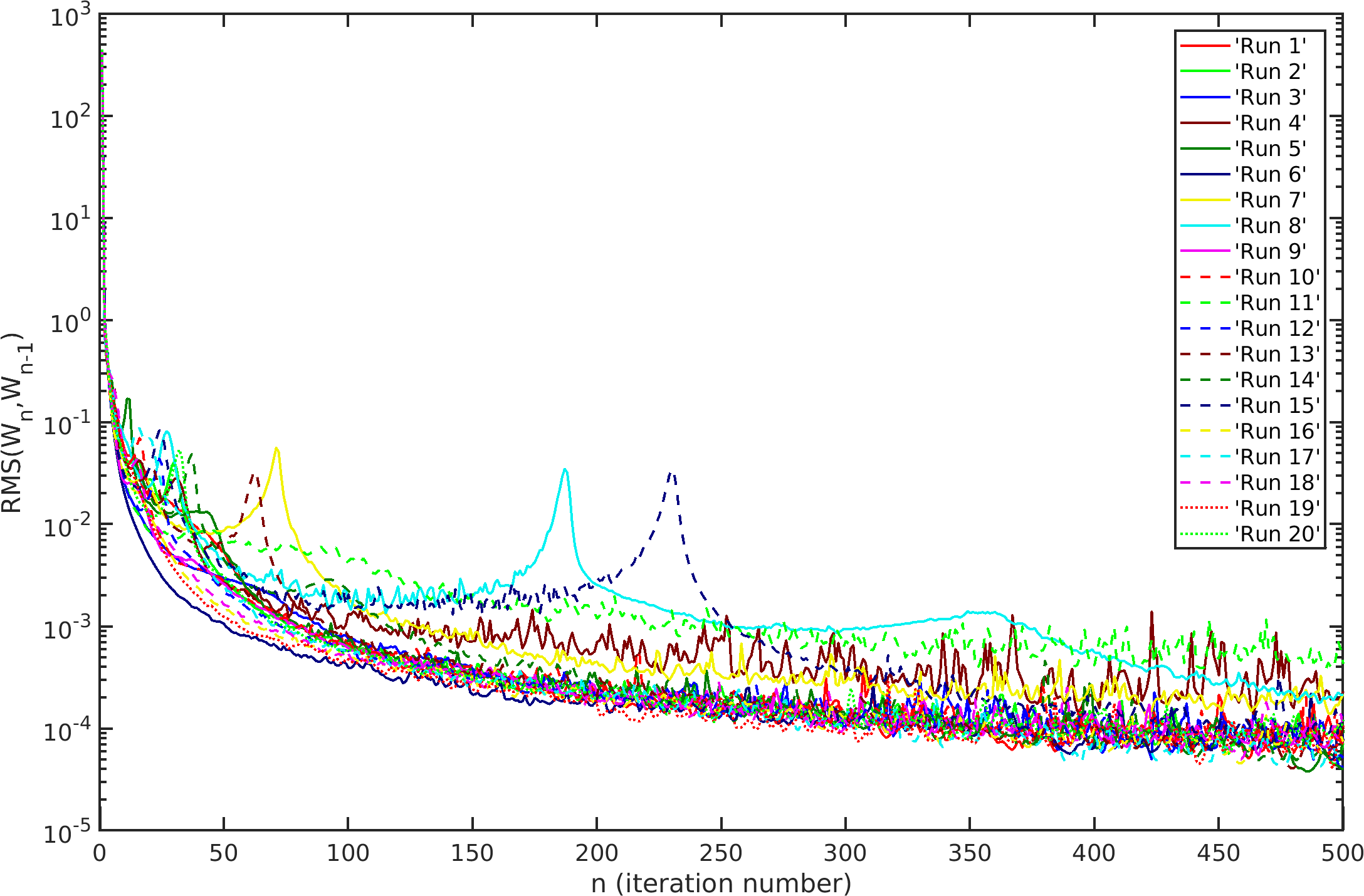}
    \caption{Root-mean-square change between neighboring iterates of Fourier space intensity $W(\mathbf{q})$ for 20 independent runs of EMC over 500 iterations each.}
    \label{fg:RMS}
\end{figure}
\begin{figure}
    \centering
    \includegraphics[width=\columnwidth]{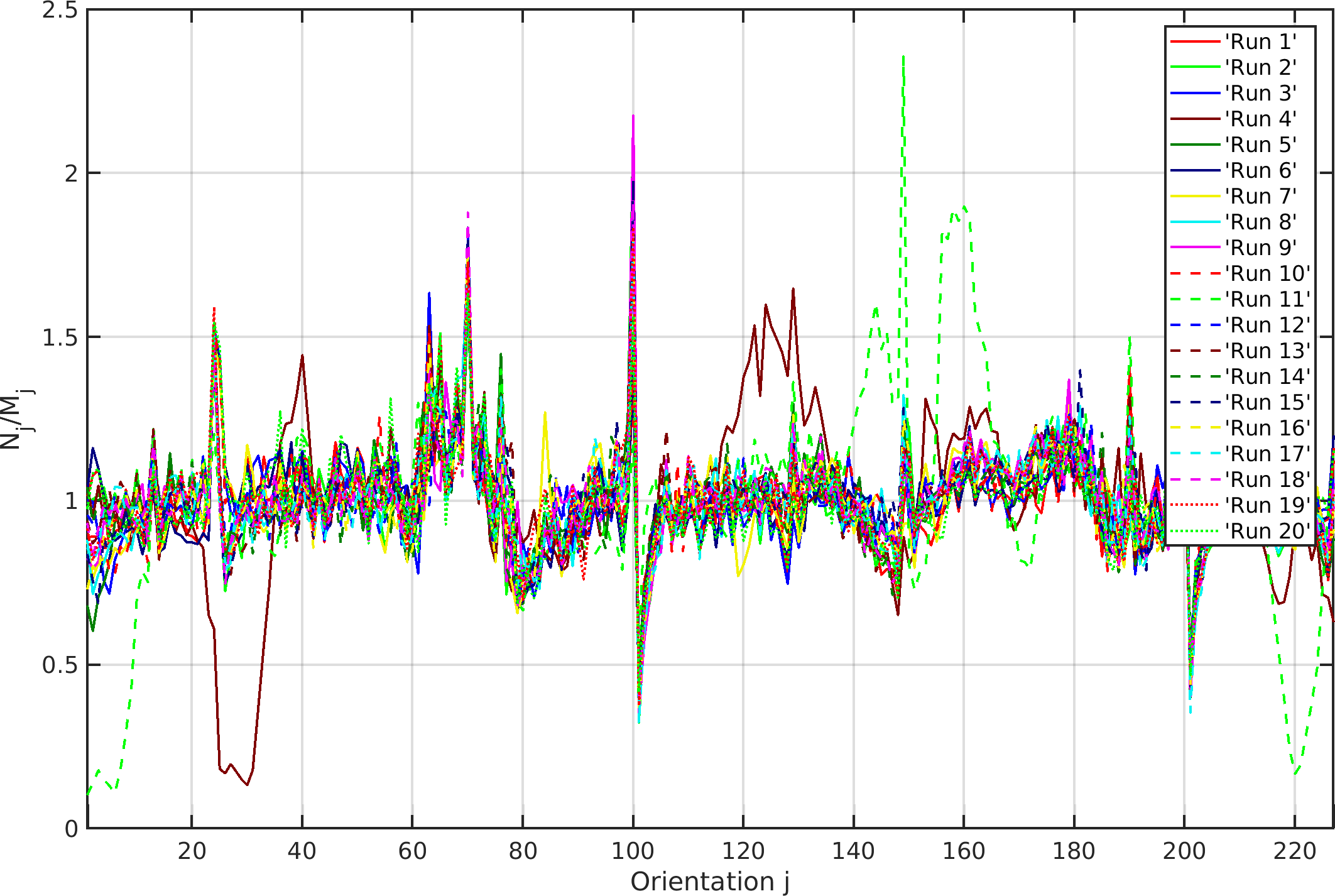}
    \caption{Distributions of orientational occupancies, normalized by the constant expected occupancy number for each orientation.}
    \label{fg:occupancies}
\end{figure}
\begin{figure}
    \centering
    \includegraphics[width=\columnwidth]{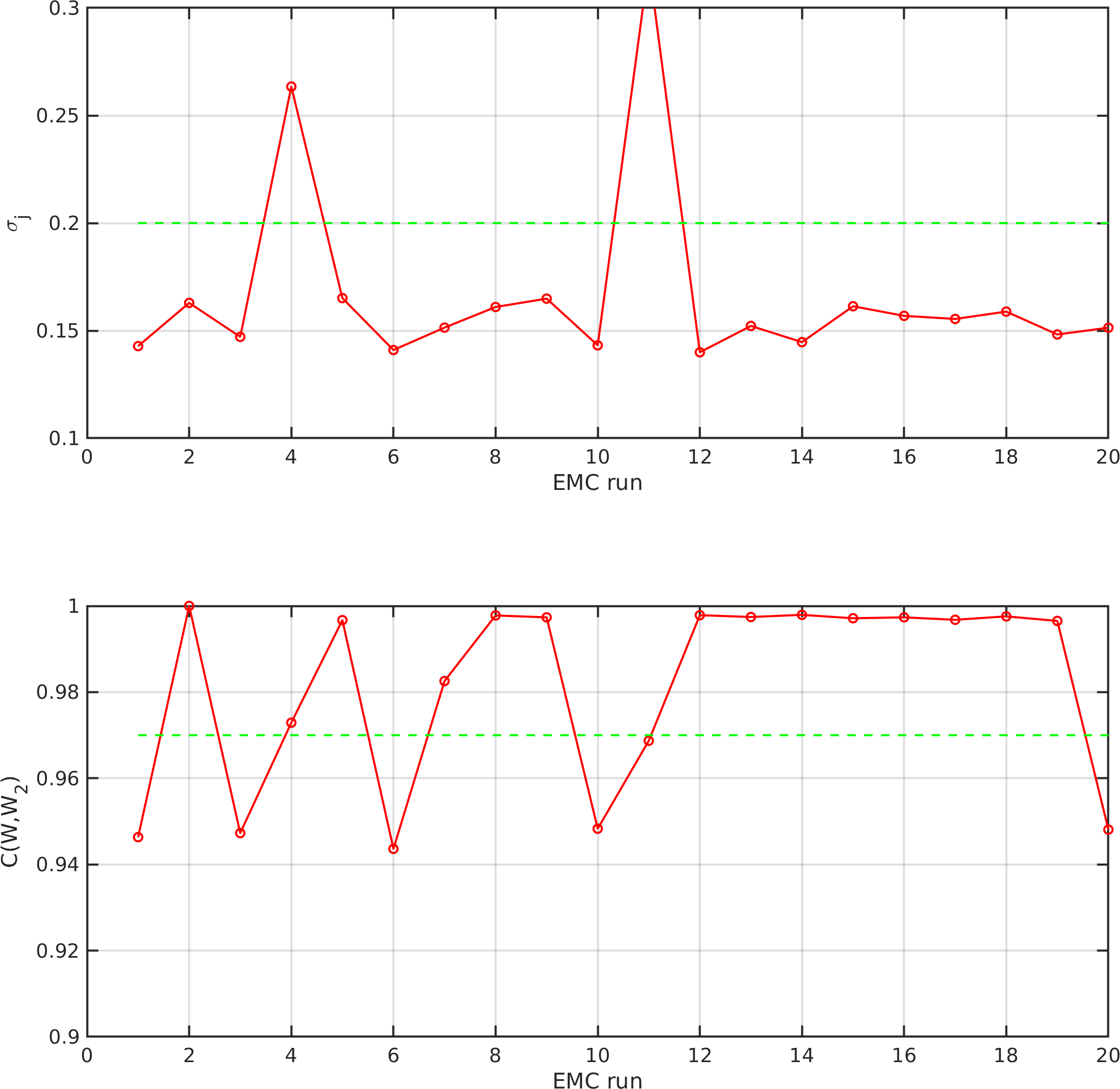}
    \caption{(top) Standard deviation of normalized occupancy numbers for each run. (bottom) Pearson correlation of obtained Fourier space distributions with a pre-selected reference distribution. Here, the reference distribution was chosen to be the result of run 2.}
    \label{fg:selection}
\end{figure}

Convergence of EMC was monitored using the root-mean-square (RMS) deviation between subsequent iterates of the Fourier space intensity $W(\mathbf{q})$ \cite{loh_reconstruction_2009}. RMS curves for 20 independent runs of EMC for 500 iterations each are shown in Fig.\ \ref{fg:RMS}: The majority of runs show a rapid decay of RMS values into a nearly steady state after an initial local maximum. However, there are exceptions, where a distinct local maximum develops at higher iteration numbers.

To arrive at a figure of merit for the quality of the converged result the occupancies $N_j$ of orientations in the last iteration of EMC were investigated. In order to calculate $N_j$, the most likely orientation for a given data frame as obtained by EMC was used, even though the algorithm itself forms each slice $W_j$ as a sum of all data frames, weighted by their orientational probabilities. Fig.\ \ref{fg:occupancies} shows the distributions of orientational occupancies $N_j$, normalized by the constant total number $M_j$ of frames in each orientation ($M_j = 2000$). As the data frames are experimentally equally distributed over all orientations, an equal orientational distribution is the ideal result. For the given dataset, a clear separation between likely and non-likely solutions can be observed, based on the occupancy standard deviation $\sigma(N_j/M_j)$ over orientations $j$, as shown in Fig.\ \ref{fg:selection}(top). In conclusion, all results with a standard deviation $\sigma(N_j/M_j) > 0.2$ were excluded from further analysis. In the present example, these were two out of 20 EMC reconstructions.

In a second step, results were classified according to their overall orientation. To this end, orthogonal slices of the obtained intensity distributions were compared visually with orthogonal slices of the intensity distribution $W_\mathrm{ref}$, which had been manually assembled using known orientations of data frames. Two classes could be identified, one of which (`class 1') exhibited strongest similarity with the reference distribution. This visual classification was accompanied by correlating results with a pre-selected reference result. To make this comparison independent of the manually obtained reference distribution which is generally not available in an FEL-based single particle imaging experiment, a representative example of class 1 was chosen as a reference for all other results. Then, the Pearson correlation coefficient was calculated including voxels within a shell ($20 \leq q \leq 50$) in Fourier space where the differences between the two classes were most prominent. The resulting correlation values can also be separated into two groups, based on a threshold of $0.97$, as shown in Fig.\ \ref{fg:selection} (bottom).

\begin{figure}
    \centering
    \includegraphics[width=0.9\columnwidth]{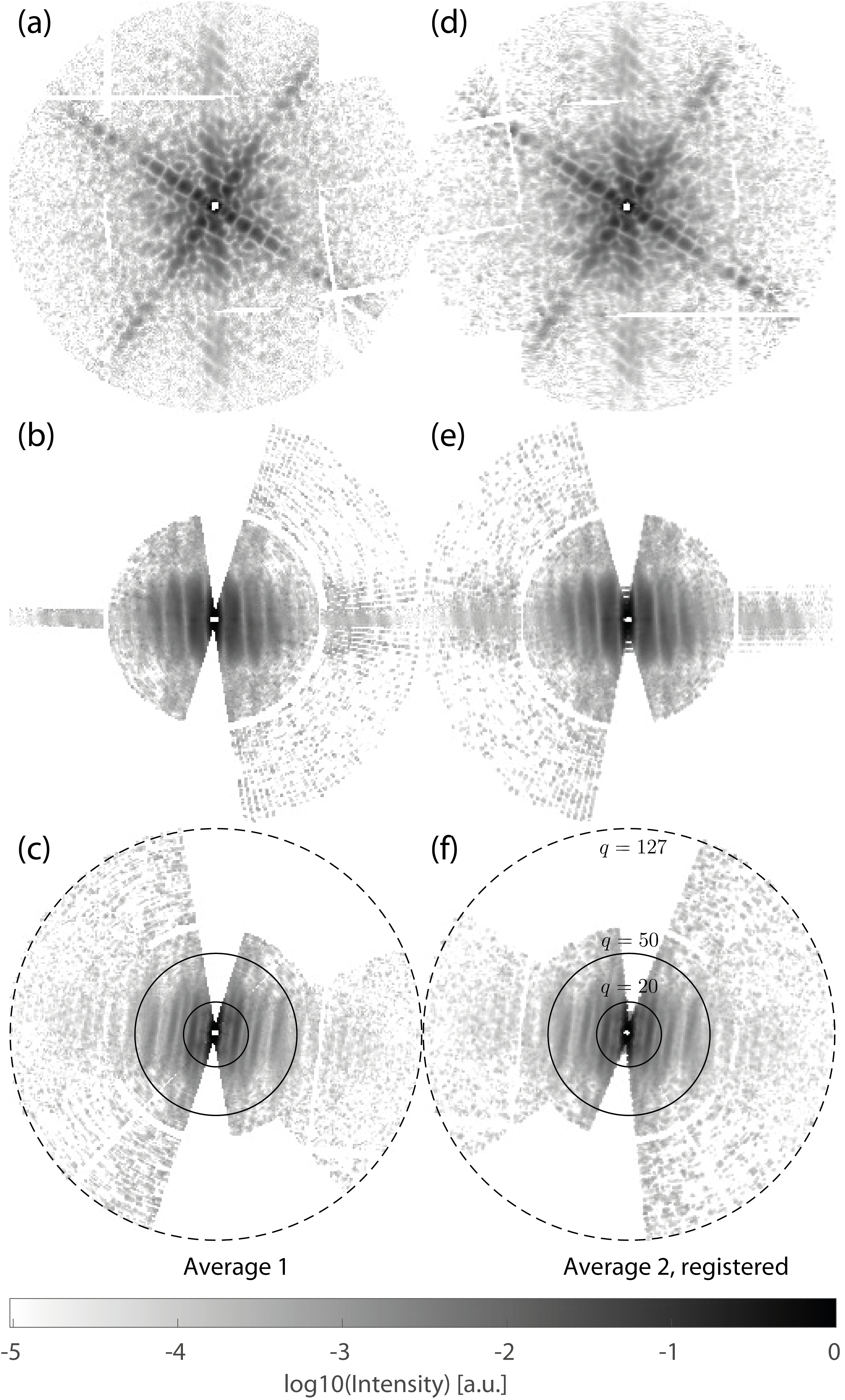}
    \caption{Orthogonal slices through Fourier space intensities obtained from averaging over two classes of the obtained results (see SM text). The result of class 1 (`Average 1'; a, b, c), also shown in Fig.\ 2 of the main text, is compared here to the rotationally registered result of class 2 (`Average 2'; d, e, f), verifying their close similarity.}
    \label{fg:registration}
\end{figure}

The resulting intensity distributions corresponding to the two classes of results are illustrated in Fig.\ \ref{fg:registration}. Shown here are distributions resulting from averaging over 13 (`Average 1') and 5 (`Average 2') out of 20 intensity distributions. To verify that the relation between the two classes of results is an overall rotation about an axis close to the $z$-axis, the two Fourier space volumes were registered with respect to each other using the same principle as described in Section \ref{sc:alignment}. After registration, the two distributions exhibit a correlation value of $99.4\%$ within the Fourier space shell $20\leq q\leq 50$.

\subsection{FRC determination}

As described in the main text, the dataset was randomly split into two halves in order to obtain a self-consistent criterion for the validity of the reconstructed reciprocal space volume. EMC was then applied two both half-datasets as described in the previous section. Notably, with only 50\% of the frames in each dataset, a significant increase in the number of unsuccessful intensity reconstructions was observed, as indicated by the standard deviation of occupancy distributions and by artifacts in the reconstructed intensity distributions. From the 20 reconstructions performed for each half 6 were disregarded for the first half and 4 were disregarded for the second half. The applied threshold was the same as before.

The ratio of reconstructions belonging to class 1 and 2 was similar to the whole dataset in both cases (9 vs.\ 5 and 10 vs.\ 6, respectively). It is noted that the separation of both orientational classes was not as obvious as for the whole dataset, indicating that $M_{data}$ should not be reduced much below the experimental value for the given dataset.

\subsection{Real space reconstruction (phase retrieval)}

To enable convergence, pixels dominated by signal from the direct beam near the center of the 3D diffraction volume were masked out before iterative phasing. As an initial guess for the iterative reconstruction, a sphere with a radius of 40 pixels filled with random values was used. During the 420 HIO iterations the object support was gradually refined using the shrinkwrap algorithm applied at every 20th iteration \cite{marchesini_x-ray_2003}. Here, an amplitude-threshold of 10\% of the maximum was applied to update the support mask, after application of a Gaussian low-pass filter with $\sigma_i = 1/8 + 7/8\cdot\exp(-i/N_\mathrm{SW})$ for iteration number $i = 1,2,\dots,420$ and $N_\mathrm{SW} = 420$. 

\section{Summary of essential experimental and analysis parameters}

A summary of essential experimental and analysis parameters of the present experiment is given in Table \ref{tb:summary}. The values are given here to allow for a detailed comparison to previous (and future) studies on simulated and experimental datasets of Single Particle Coherent Diffractive Imaging. All definitions are in accordance with Ref.~\onlinecite{loh_reconstruction_2009}, unless otherwise noted in the main text. As the sample particle in the present study is non-spherical, its dimensionless particle radius is given here along the direction of the symmetry axes of the smallest rectangular cuboid containing the particle. The linear sampling ratio is given along the same directions. The reduced information rate $r(N) = \langle I(K,W)/I(K,W)|\Omega\rangle$ \cite{loh_reconstruction_2009} was averaged here over the last 300 iterations of each run and then averaged over the 13 runs that were used to obtain $W_\mathrm{main}$ (see main text). $\delta\theta$ defines the minimum angle between two orientations that is at least required to adequately sample the 3D intensity distribution. $R_\mathrm{max}$ here denotes the radius of the particle in resolution elements along the direction of its largest extension.

\section{Significance of the observed level of sparsity}
\label{sc:sparsity}

\balance

\begin{figure}
    \centering
    \includegraphics[width=1.1\columnwidth]{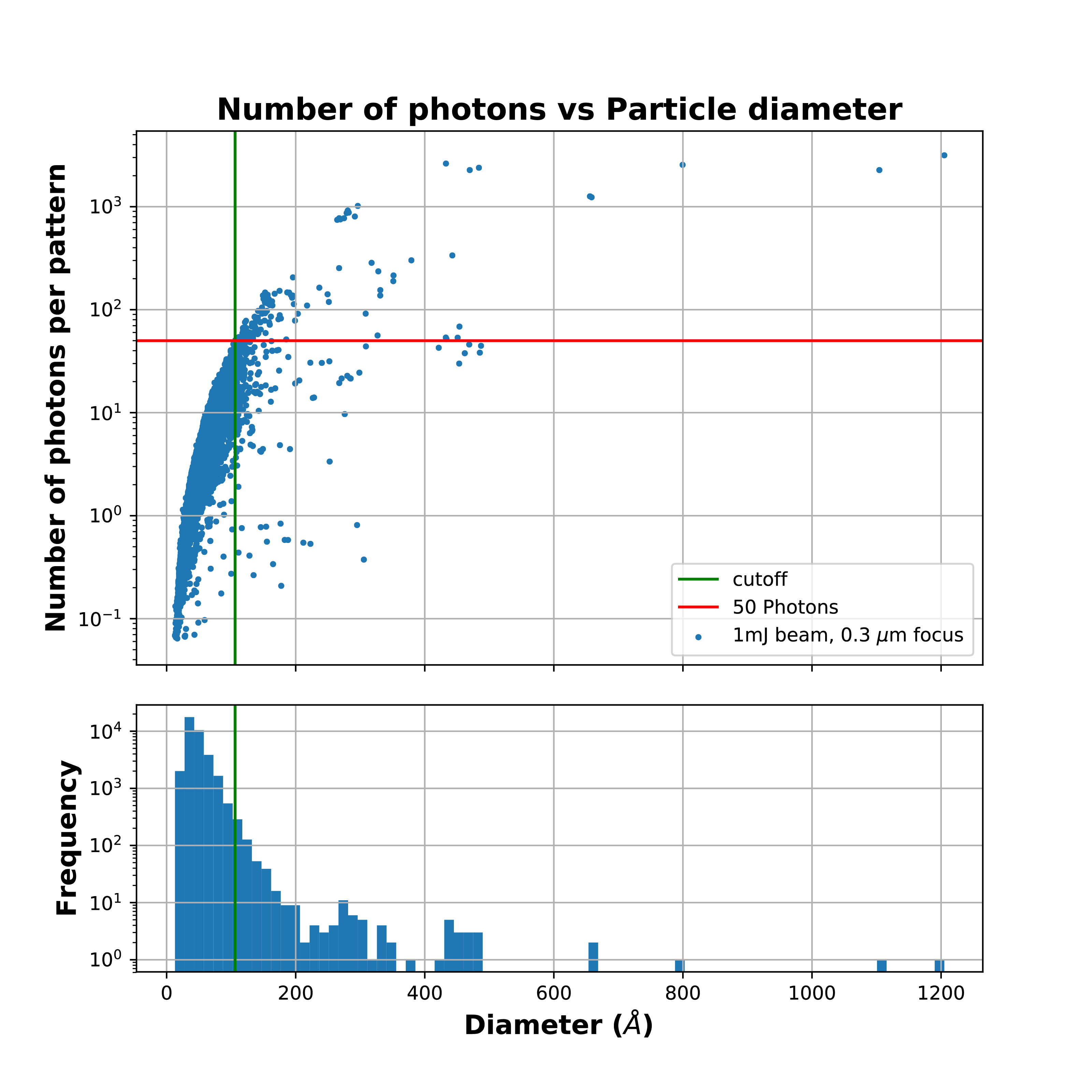}
    \caption{(Top) The number of scattered photons outside the central speckle is plotted for 35000 human proteins from the PDB against their diameter, assuming an incident pulse energy of \unit{1}{\milli\joule} and a focus diameter of \unit{300}{\nano\metre} at \unit{8.1}{\kilo\electronvolt} photon energy. The red horizontal line indicates a level of 50 scattered photons. The green vertical separates particles which scatter less at given conditions (left) and more (right). The separation is at a particle diameter of \unit{10.6}{\nano\metre}. (Bottom) Size distribution of the 35000 protein structures selected from the PDB.}
    \label{fg:nphot}
\end{figure}

The following assumptions were used to estimate the average particle size that elastically scatters a target number of photons (here: 50) per pattern. 
\begin{enumerate}
\item{} The number of X-ray photons detected by a detector pixel is assumed to be
\begin{equation}
I(\mathbf{Q_{\rm pix}}) = I_0 r_e^2 \delta \Omega_{\rm pix} |F(\mathbf{Q_{\rm pix}})|^2, \label{eqn:ScatteringEqn}
\end{equation}
where $r_e$ is the classical electron radius, $I_0$ is the X-ray fluence, $\delta \Omega_{\rm pix}$ is the solid angle subtended by the pixel, and $|F(\mathbf{Q_{\rm pix}})|^2$ is the elastic scattering factor of the particle averaged over all possible particle orientations. $\mathbf{Q_{\rm pix}}$ here denotes the unit-carrying reciprocal space-coordinate of a detector pixel. We estimate this scattering factor with the Debye scattering equation; here all pixels with a common distance to the central beam on the detector have the same modulus $Q = |\mathbf{Q_{\rm pix}}|$:
\begin{equation}
	|F(Q)|^2 = \sum_{i,j}^{N_{\text{atoms}}} f_i f_j \frac{\sin Qr_{ij}}{q r_{ij}}.
\end{equation}
This scattering factor was computed factor for approximately 35,000 human protein structures deposited in the RCSB Protein Data Bank, where the atomic scattering factors ($f_i$) were taken from Ref.~\onlinecite{henke_x-ray_1993}. Since we are only targeting resolutions coarser than \unit{5}{\angstrom} here, the atomic form factors are ignored in these calculations. Further, these proteins were assumed to be in vacuum and not solvated during these scattering simulations.

\item{} The photon energy has assumed to be \unit{8.1}{\kilo\electronvolt}. The focus diameter and pulse energy has been varied within $0.1$ to \unit{0.3}{\micro\metre} and $0.1$ to \unit{1}{\milli\joule}, respectively (see Table \ref{table:ProteinSizes}).

\item{} For simplicity, we simulated photon counts on circular detectors with a fixed maximum resolution of \unit{3}{\angstrom}, but different beamstops for different proteins such that each beamstop spans the central speckle for each protein structure. We obtain the average total number of elastically scattered photons for each protein by integrating Eqn. \eqref{eqn:ScatteringEqn} azimuthally across the pixels on this detector. 

\item{} The average total photons from the previous step are computed for all 35000 protein structures and scaled to account for different pulse focus diameters. Table \ref{table:ProteinSizes} shows the diameter of the smallest protein that scatters at least 50 photons across the detector described above. 
\end{enumerate}

The number of scattered photons for any selected particle from the PDB, as simulated for the case of \unit{1}{\milli\joule} pulse energy and \unit{0.3}{\micro\metre} focus diameter is shown in Fig.\ \ref{fg:nphot}. It can be seen that hundreds out of the 35000 protein structures scatter more than 50 photons, namely those with a diameter larger than about \unit{10}{\nano\metre}. However, the majority of proteins from the given ensemble has a smaller diameter and scatters even less, under the given conditions. 

This shows that the number of scattered photons as observed in the present experiment provides a very realistic test case for an FEL-based SPI experiment from a relevant protein structure. 

\section{Tables}

\begin{table}[!h]
\centering
\begin{tabular}{l | l}
\hline
\textbf{Parameter} & \textbf{Value} \\
\hline
$R$ (particle radius in resolution elements)  & \vtop{\hbox{\strut ca. $2.5\dots26$,}\hbox{\strut depends on}\hbox{\strut direction}}\\
$N$ (mean number of photons per pattern) & $49.3$ \\
$\Tilde{N}$ (median number of photons per pattern) & $48$ \\
\vtop{\hbox{\strut $\sigma(N)$ (standard deviation}\hbox{\strut of photons per pattern)}}& $9.6$\\
$\langle r(N)\rangle$ (mean reduced information rate) & $0.86$\\ 
\vtop{\hbox{\strut $\delta\theta=1/R_\mathrm{max}$ (required angular}\hbox{\strut scale of orientations)}} & $\unit{2.2}{\degree} $  \\
$M_{data}$ (total number of data frames) & 454,000  \\
$M_{rot}$ (total number of unique orientations) & 227 \\
\hline
\end{tabular}
\caption{\label{tb:summary}Main parameters of the experiment. As the particle is flat rather than spherical, 
some parameters vary within a certain range, rather than being constrained to a single value. Minima and maxima here correspond to directions along the particle sides. For further details, see the text of the appendix.}
\end{table}

\begin{table}[!h]
\centering
\begin{tabular}{c | c  c  c  c}
\hline
\diagbox[width = 0.23 \textwidth]{\vspace*{0.1cm}\small \textbf{focus diameter}}{\vspace*{0.1cm}\small \textbf{pulse energy}} & \unit{1}{\milli\joule} & \unit{0.5}{\milli\joule}& \unit{0.3}{\milli\joule}& \unit{0.1}{\milli\joule} \\
\hline
 \unit{0.3}{\micro\metre} & 106 & 148 & 196 & 264 \\
 \unit{0.1}{\micro\metre} & 51.2 & 63.6 & 74.1 & 117  \\
\hline
\end{tabular}
\caption{Particle diameter (in \unit{\!\!\!}{\angstrom}) of the smallest protein that scatters at least 50 photons per detector pattern when averaged over all orientations, at various pulse energies and focus diameters. This table is computed from more than 35,000 human protein structures in the RCSB Protein DataBank (PDB).}\label{table:ProteinSizes}
\end{table}



\acknowledgments
Detector research at Cornell is supported by U.S. Department of Energy awards DE-SC0004079, DE-SC0016035 and DE‐SC0017631 a by the Cornell High Energy Synchrotron Source (CHESS), supported by the U.S. National Science Foundation (DMR-1332208). This research used resources of the National Synchrotron Light Source II, a U.S. Department of Energy (DOE) Office of Science User Facility operated for the DOE Office of Science by Brookhaven National Laboratory under Contract No. DE-SC0012704. The experiments were performed on beamline ID10 at the European Synchrotron Radiation Facility (ESRF), Grenoble, France. The work was done partially while K.G. was visiting the Institute for Mathematical Sciences, National University of Singapore, in 2018. The visit was supported by the Institute. Duane Loh and Colin Teo would like to acknowledge funding support of the Singapore
National Research Foundation’s Competitive Research Program funding (NRF-CRP16-2015-05).

Jerome Kieffer and Pierre Paleo are acknowledged for GPU-implementation of the phase iterative algorithm. The authors are grateful to Pascal Dideron, Pascal Voisin and Laurent Claustre for technical support during the experiment. 


%

\end{document}